\documentclass[12pt]{iopart}
\expandafter\let\csname equation*\endcsname\relax
\expandafter\let\csname endequation*\endcsname\relax
\usepackage{hyperref}
\usepackage{amsmath}
\usepackage{braket}
\usepackage{bm}
\usepackage{graphicx}
\usepackage{float} 
\usepackage{subfigure} 
\usepackage{color}
\usepackage{ulem}
\usepackage{soul}

\usepackage[ruled,linesnumbered]{algorithm2e}

\begin{document}

\title{Inverse iteration quantum eigensolvers assisted with a continuous variable}

\author{Min-Quan He$^1$, Dan-Bo Zhang$^{2,3, \ast}$, Z. D. Wang$^{1,3, \ast}$}
\address{$^1$ Department of Physics and HKU-UCAS Joint Institute for Theoretical and Computational Physics at Hong Kong, The University of Hong Kong, Pokfulam Road, Hong Kong, China}
\address{$^2$ Guangdong Provincial Key Laboratory of Quantum Engineering and Quantum Materials, GPETR Center for Quantum Precision Measurement and SPTE, South China Normal University, Guangzhou 510006, China}
\address{$^3$ Frontier Research Institute for Physics, South China Normal University, Guangzhou 510006, China}
\address{$^{\ast}$ Authors to whom any correspondence should be addressed.}
\ead{dbzhang@m.scnu.edu.cn and zwang@hku.hk}

\begin{abstract}
    
    The capacity for solving eigenstates with a quantum computer is key for ultimately simulating physical systems. Here we propose inverse iteration quantum eigensolvers, which exploit the power of quantum computing for the classical inverse power iteration method. A key ingredient is constructing an inverse Hamiltonian as a linear combination of coherent Hamiltonian evolution. We first consider a continuous-variable quantum mode~(qumode) for realizing such a linear combination as an integral, with weights being encoded into a qumode resource state. We demonstrate the quantum algorithm with numerical simulations under finite squeezing for various physical systems, including molecules and quantum many-body models. We also discuss a hybrid quantum-classical algorithm that directly sums up Hamiltonian evolution with different durations for comparison. It is revealed that continuous-variable resources are valuable for reducing the coherent evolution time of Hamiltonians in quantum algorithms. 

\end{abstract}

\maketitle

\section{Introduction}

Solving eigenstates of many-body interacting Hamiltonian has caught a lot of attention for past decades. Among many proposed quantum algorithms, two large groups are quantum phase estimation (QPE)~\cite{kitaev1995quantum} and variational quantum eigensolver (VQE)~\cite{peruzzo2014variational}. Remarkably, those two quantum algorithms exploit quantum resources in different strategies: QPE can estimate high accuracy eigenvalues with constant samples at the cost of large coherent circuit depth~\cite{lanyon2010towards, du2010nmr, hoffman2014mechanism, aspuru2005simulated, wiebe2016efficient}; VQE can efficiently reduce the requirement of hardware coherent time with specified wavefunction ansatz but at a price of increasing measurement repetitions for estimating observable~\cite{peruzzo2014variational, yung2014transistor, mcclean2016theory, o2016scalable, zhang2020collective}. With but not limited to quantum eigensolvers as examples, a trade-off between available quantum resources becomes an important theme in quantum algorithm designs.

The inverse power iteration~(IPI) method is a standard numeral tool for solving eigenstates of quantum systems~\cite{pohlhausen1921berechnung}. It is not scalable for large quantum systems, as the time complexity of inverting Hamiltonian grows exponentially with system size due to the exponential growth of the dimension of Hilbert space~\cite{le2014powers}. Since a key ingredient is to perform an inverse operation on a Hamiltonian, it is natural to incorporate many developed quantum algorithms for matrix inversion~\cite{harrow2009quantum,clader2013preconditioned, rebentrost2014quantum, schuld2016prediction, kerenidis2016quantum, wang2017quantum} to endow quantum advantages for the IPI method. Nevertheless, the inverse matrix is generally non-unitary, and its construction can be rather resource-consuming. By expressing inverted Hamiltonian as an integral of Hamiltonian evolutions, a hybrid quantum-classical algorithm has been proposed recently using a summation of Hamiltonian dynamics at discrete time~\cite{kyriienko2020quantum}, without referring to auxiliary qubits or qumodes and thus more feasible on near-term quantum devices. On the other hand, implementing quantum matrix inversion can be simplified with continuous-variable quantum modes~(qumode), which naturally implements integral of unitaries without discretization~\cite{lau2017quantum, arrazola2019quantum, zhang2019realizing, zhang2020protocol}. Along this line, we pursue a further simplified quantum algorithm with only one qumode to invert Hamiltonian. This makes the full quantum version of IPI method easier to implement, given that manipulation of qumode state and its coupling to qubits are highly controllable in some mainstream quantum platforms, e.g., superconducting circuits~\cite{wallraff2004strong, paik2011observation, devoret2013superconducting} and trapped ions~\cite{leibfried2003quantum, monroe2013scaling, gan2020hybrid}. Moreover, the valuable continuous-variable resource for solving eigenstates as well as other computational tasks may be revealed~\cite{liu2016power}.

In this paper, we propose a quantum inverse power iteration~(QuIPI) algorithm with a single ancillary qumode for solving the ground state and other eigenstates of quantum systems. The QuIPI involves inverting Hamiltonian as a key subroutine, which is constructed with a linear combination of unitaries, with weights being encoded in a specified qumode resource state. We present a decomposition scheme of time evolution for coupled system-qumode Hamiltonian into basic quantum gates of qubits and one qubit-qumode gate. The time complexity is a polynomial of the system size and a quasi-polynomial of the desired accuracy. We numerically simulate the quantum algorithm for models ranging from quantum chemistry to quantum many-body systems, including molecular hydrogen, quantum Ising model, and Kitaev ring. We also discuss a hybrid quantum-classical algorithm of IPI, where only evolution of system Hamiltonian with different evolution time is required. A comparison to QuIPI is made to stress the role of continuous-variable resources for reducing the requirement of long-time evolution of the quantum system. Typically, the qubit-based method such as the HHL algorithm~\cite{harrow2009quantum} for matrix inversion requires a large number of ancillary qubits. Moreover, it relies on a quantum Fourier transformation which involves a complicated quantum circuit. On the other hand, the QuIPI exploits only a single ancillary qumode which can greatly simplify the algorithm. Furthermore, compared to the existing qumode-assisted matrix inversion approaches~\cite{zhang2019realizing, zhang2020protocol}, the QuIPI is easier to implement since it only requires entangling a single qumode to qubits.

This paper is organized as follows. We firstly present the quantum algorithm in Sec.~\ref{Sec Quantum eigensolver}. Then numerical results of solving ground state energy are shown in Sec.~\ref{Sec Numerical result}. Finally, conclusions and discussions are given in Sec.~\ref{sec Discussion and conclusion}.

\section{Quantum eigensolver}
\label{Sec Quantum eigensolver}
In this section, we formulate the QuIPI eigensolver. We first introduce continuous-variable assisted quantum algorithms and show how they can be used for the inverse power iteration method. Then we will present the detailed quantum algorithm, including initial state preparation, unitary evolution, and projection. Lastly, the time complexity of this algorithm will be analyzed.

\subsection{Continuous-variable assisted quantum algorithm}
\label{One qumode assisted quantum algorithm}
A quantum state with one continuous variable can be written as $\int_{-\infty}^{\infty} \psi(p) \ket{p} dp$ or its conjugate one $\int_{-\infty}^{\infty} \phi(q) \ket{q} dq$, where $\ket{p}$ and $\ket{q}$ are eigenstates of two conjugate quardatures~(such as momentum and position) $\hat{p}$ and $\hat{q}$. Unlike a qubit being a two-level state; a qumode has infinite dimensionality. Moreover, as $[\hat{p}, \hat{q}] = i$, there is an intrinsic Fourier transformation, $\ket{q} = \frac{1}{\sqrt{2 \pi}} \int_{-\infty}^{\infty} e^{-iqp} \ket{p} dp$. Those properties make continuous variables capable of encoding and processing high-density information which are valuable for quantum computing. 

The quantum computing model of continuous-variable assisted quantum algorithm encodes the system with qubits and refers to a few qumodes as ancilla. Such a hybrid-variable approach of quantum computing consisting of both qubits and qumodes has a potential advantage to make the best-of-both-worlds~\cite{furusawa2011quantum,andersen_15,liu2016power,gan2020hybrid}. It has been applied for different areas, e.g., solving linear partial differential problems~\cite{arrazola2019quantum}, quantum machine learning~\cite{zhang2019realizing, zhang2020protocol}, and simulation of finite-temperature quantum systems~\cite{zhang2020continuous}. Qumodes in those algorithms are prepared in a desirable resource state, evolved by coupling to qubits, and finally projected onto a given qumode state. The  ancillary role reveals that the qumodes are decoupled from the system of qubits at the final stage. 
 
The continuous-variable assisted quantum computing is similar to the conventional approach with qubits, but with additional requirements for well-control of the qumodes and their coupling to qubits. Firstly, manipulation of qumodes for generating complicated continuous-variable quantum states has been sufficiently advanced with the current optimal quantum control techniques~\cite{heeres_implementing_2017,ma_error-transparent_2020}. Secondly, unitary operators of the coupling between qubits and qumodes can be universally decomposed into basis quantum gates of qubits and one hybrid qubit-qumode gate~\cite{zhang2020protocol,zhang2020continuous}(also see~\ref{Arbitrary qubit-qumode local operator construction}). Moreover, the continuous-variable assisted quantum algorithm is physically realizable, by exploiting continuous variables naturally existing in the mainstream platforms, such as motional modes~(phonons) of trapped ions~\cite{leibfried2003quantum, monroe2013scaling, gan2020hybrid} and cavity modes~(photons) of superconducting circuit systems~\cite{wallraff2004strong, paik2011observation, devoret2013superconducting}.

\subsection{Quantum version of inverse power iteration method}
\label{Quantum version of inverse power iteration method}
In the IPI method, the ground state of a given Hamiltonian $\hat{H}$ is determined by iteratively performing inverse Hamiltonian $\hat{H}^{-1}$ on an initial state $\ket{b}^{(0)}$ with prior knowledge. The iteration is shown as
\begin{equation}
    \ket{b}^{(k+1)} = \frac{\hat{H}^{-1} \ket{b}^{(k)}}{||\hat{H}^{-1} \ket{b}^{(k)}||}.
\end{equation}
Here, a shift of energy is applied to keep all the eigenvalues positive to ensure the final state converges to an approximate ground state $\ket{\psi_{g^{\prime}}}$ after $K$ steps. We assume a local Hamiltonian $\hat{H} = \sum_{l=1}^{L} c_{l} \hat{h}_{l}$, where $\hat{h}_{l}$ is a tensor product of pauli matrices $\hat{h}_{l} = \otimes_{i=1}^{N} \hat{\sigma}^{(i)}$ with $\hat{\sigma}^{(i)} \in \left\{ \hat{\sigma}_{x}, \hat{\sigma}_{y}, \hat{\sigma}_{z}, \hat{I} \right\}$, $c_{l}$ is the corresponding coefficient, $N$ is the number of qubits, and $L$ is a polynomial of system size.

To realize the non-unitary inverse-power operator $\hat{H}^{-1}$ on a quantum computer, we adopt the methodology of integral-of-unitaries~(similar to linear-combinations-of-unitaries), which can be implemented with an ancillary qumode~\cite{lau2017quantum, arrazola2019quantum,zhang2020protocol,zhang2019realizing}. Using $a^{-1} = i\int_{0}^{\infty} e^{-i a b} db$ from Fourier transformation, the inverse Hamiltonian can be expressed as
\begin{equation}
    \hat{H}^{-1} = i \int_{0}^{\infty} e^{-i \hat{H} p} dp
    \label{Integral of unitary operator}
\end{equation}

Ideally, such a non-unitary operator can be obtained by performing a unitary operator $\hat{U} = e^{-i \hat{H} \hat{p}}$ on both qubits $\ket{b}$ and a resource state ancillary qumode $\ket{R} = \int_{0}^{\infty} \ket{p} dp$ (not normalizable), then projecting the ancillary qumode on zero position state $\ket{q=0} = \int_{-\infty}^{\infty} \ket{p} dp$ (not normalizable). The result state $\ket{\psi^{\prime}}$ (not normalizable) is shown as
\begin{equation}
    \label{result state}
    \begin{aligned}
        \ket{\psi^{\prime}} = -i \hat{H}^{-1} \ket{b} = \int_{0}^{\infty} e^{-i \hat{H} p} \ket{b} dp \propto \braket{q=0|\hat{U}|R}\ket{b}.
    \end{aligned}
\end{equation}
Here both the resource state and the projection state are infinitely squeezed, which are not physical since they need infinite energy to prepare. Moreover, projecting onto an infinitely squeezed state typically leads to a zero success rate. We use the ideal case with infinite squeezing to illustrate the main idea for constructing inverse Hamiltonian, but finite squeezed states are adopted in the rest of the paper~(detail is shown in~\ref{Finite squeezed result}).

After $K$ step iterations of above process, the final state converges to the approximate ground state $\ket{\psi_{g^{\prime}}}$. Then the ground state energy $E_{0}$ is estimated by
\begin{equation}
    \braket{\psi_{g^{\prime}}|\hat{H}|\psi_{g^{\prime}}} = E_{0} + O(|b_{0}|^{2} \cdot (E_{0}/E_{1})^{2K})
    \label{Energy estimation},
\end{equation}
where $|b_{0}|^{2}$ is the overlap between the initial state and the exact ground state, and $E_0$ and $E_1$ are ground state energy and first excited state energy, respectively. This procedure can be completed by quantum expectation estimation (QEE)~\cite{peruzzo2014variational}. Eq.~\eqref{Energy estimation} shows the energy error exponentially decays with iteration step $K$ (see~\ref{Energy error result from inverse iteration} for more details).

\subsection{Procedure of quantum algorithm}
\label{Procedure of quantum algorithm}
\begin{figure}
    \centering
    \includegraphics[width=0.48\textwidth]{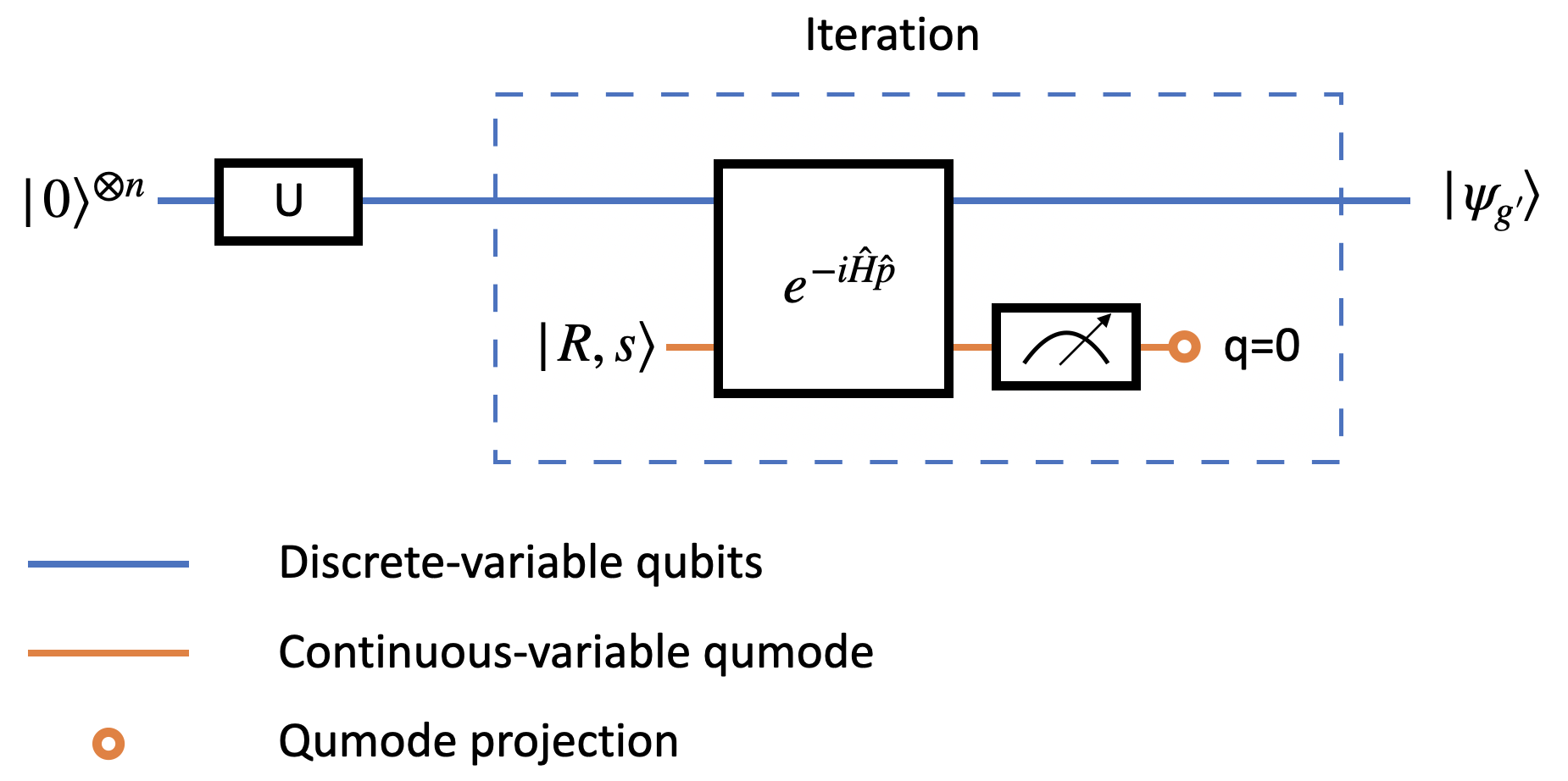}
    \caption{Illustration of preparing the ground state for a Hamiltonian $H$ with the QuIPI algorithm.}
    \label{Quantum circuit}
\end{figure}

We now present the algorithmic procedure of QuIPI for solving the ground state and estimating the ground state energy. The quantum circuit is shown in Fig.~\ref{Quantum circuit} and involves the following steps:
\begin{itemize}
    \item[1.] \textbf{State preparation.}  
    The qubits are prepared in a state with a nonzero overlapping with the target ground state. The qumode is initialized in a finite squeezed resource state $\ket{R,s} = \sqrt{2} s^{-1/2} \pi^{-1/4} \int_{0}^{\infty} e^{-p^2/2s^2} \ket{p} dp$. By writing $\ket{R,s}$ in the Fock state space with a cutoff of Fock number, we can efficiently prepare the resource state, which is discussed in the~\ref{Demonstration of resource state preparation}.
    
    \item[2.] \textbf{Unitary operator performing.} Performing a unitary operator $e^{-i\hat{H}\hat{p}}$ on both qubits and the ancillary qumode. The unitary operator can be decomposed as a set of universal single-qubit and two-qubit quantum gates and a qubit-qumode coupling gate $e^{-i\sigma^{x} \hat{p}}$, which will be given later. 
    
    \item[3.] \textbf{Projection.} Projecting the ancillary qumode on the finite squeezed position state $\ket{q,s} = s^{-1/2} \pi^{-1/4} \int_{-\infty}^{\infty} e^{-p^2/2s^2} \ket{p} dp$, the final state at large $s$ limit turns to be, 
    \begin{equation}\label{eq:final_state}
    \ket{\psi^{\prime}} = - i \sqrt{\frac{2}{\pi}} s^{-1} \sum_{n}( E_{n}^{-1} + O(s^{-2})) \ket{\psi_{n}}.
    \end{equation}
    When $s\rightarrow\infty$, it recovers to the ideal case in Eq.~\ref{result state}. 
    For the general case, a concrete expression is given in~\ref{Finite squeezed result}.
\end{itemize}

The above procedure should be repeated K times. In each iteration, only if the ancillary qumode is successfully projected on the target state, the procedure continues; and remarkably, the qumode is re-prepared in the resource state $\ket{R,s}$.

After successfully preparing the approximate ground state $\ket{\psi_{g^{\prime}}}$, the last step is to estimate the ground state energy by QEE~\cite{peruzzo2014variational}, in which the expectation value of Hamiltonian is decomposed into several expectations of local operators,
\begin{equation}
    E_{0}^{\prime} = \braket{\psi_{g^{\prime}}|\hat{H}|\psi_{g^{\prime}}} = \sum_{l=1}^{L} c_{l} \braket{\psi_{g^{\prime}}| \hat{h}_{l} |\psi_{g^{\prime}}}.
\end{equation}
We individually measure the expectation value of each local operator $\braket{\psi_{g^{\prime}}|\hat{h}_{l}|\psi_{g^{\prime}}}$ by local measurements of each qubit~\cite{nielsen2002quantum}, then weighted sum up to obtain the ground state energy $E_{0}$.

We stress some physical aspects of the qumode and discuss consequences on the performance of QuIPI. In physical systems, the qumode state may only be approximately prepared.  The resource state $\ket{R,s}$ can be approximated as a superposition of Fock state with a truncation. Such a superposed state can be well-prepared for a moderate truncation, which can be sufficient for the QuIPI algorithm~(see more details in Appendix~\ref{Resource state}). Another important physical resource is squeezing. The projection can be realized by homodyne detection~\cite{poyatos1996motion},  implemented by performing a squeezing operator with a squeezing factor $s$ on the qumode and then projecting onto the vacuum state $\ket{n = 0}$ (zero Fock number state). As a post-selection process, there is a factor $s^{-1}$ in Eq.~\ref{eq:final_state}, which accounts for the successful projection rate. Since the successful projection rate has an inverse relation to the squeezing factor while the state accuracy is proportional to the squeezing factor, a larger squeezing factor leads to a more accurate solution but a lower success rate. In physical realization, a higher squeezing factor requires higher energy, and thus the squeezing can be seen as a quantum resource for implementing the quantum algorithm~\cite{liu2016power}.

\subsection{Time complexity analysis}
\label{Time complexity analysis}
We briefly analyze the time complexity, which involves the circuit depth results from Suzuki-Trotter decomposition, the repetition time caused by qumode projection, and the requirement of iteration step $K$. 

Firstly, the unitary operator $e^{-i\hat{H}\hat{p}}$ is decomposed into several quantum gates on the circuit by trotter decomposition~\cite{suzuki1991general}. The unitary operator is expressed in the following form,
\begin{equation}
    e^{-i\hat{H}\hat{p}} = (\Pi_{l=1}^{L} e^{-i c_{l} \hat{h}_{l} \hat{p} /n})^{n} + O(\frac{1}{n}),
    \label{Trotter decomposition}
\end{equation}
where each element $e^{-i c_{l} \hat{H}_{l} \hat{p} /n}$ can be constructed by a single qubit-qumode operator $e^{-i\hat{\sigma}_{x}\hat{p}}$ and some qubit gates (see~\ref{Arbitrary qubit-qumode local operator construction}). The state error is $O(\frac{1}{n})$ with $n$ steps of decomposition introducing an energy error of $O(\frac{1}{n^{2}})$. For a desired state accuracy $\epsilon^{\prime}$, the total gates required are $O(L^{3}c_{max}^{2}/\epsilon^{\prime})$, which means $O(L^{3}c_{max}^{2}/ \sqrt{\epsilon})$ gates are required in each iteration for a desired energy accuracy $\epsilon$. Totally, there are $O(KL^{3}c_{max}^{2}/ \sqrt{\epsilon})$ quantum gates in the quantum circuit. Here we adopt the simplest case, the first order Suzuki-Trotter decomposition, to estimate the circuit depth. Many improved versions of gate decomposition strategies can lower the circuit depth, such as the divide and conquer approach~\cite{hadfield2018divide} and random compiler approach~\cite{campbell2019random}.

The repetition times required is also analyzed. Considering a finite squeezed factor, the state error is proportional to $s^{-2}$, as evaluated from Eq.~\ref{eq:final_state}. So, the energy error is $O(s^{-4})$. For a desired energy error $\epsilon$, the requirement of the squeezing factor is $O(\epsilon^{-1/4})$. For each step iteration in the $K$ steps, the process continues only if the ancillary qumode is successfully projected onto the target state. The success rate in each step is $O(s^{-1})$ , and the total repetition time for $K$ steps thus is $O(s^{K}) = O(\epsilon^{-K/4})$ for state preparation. Then, $O(c_{max}^{2}L\epsilon^{-2})$ times measurements are required by using QEE~\cite{peruzzo2014variational}. The number of total samples is $O(c_{max}^{2}L\epsilon^{-(2+K/4)})$. 

In total, the time complexity for solving ground state energy is $O(L^{4} c_{max}^{4} K \epsilon^{-(5/2+K/4)} )$ for requisite accuracy $\epsilon$. Recalling Eq.~\ref{estimated_energy}, the accuracy exponentially increases with the step $K$, which means the requirement of $K$ is a logarithm of the desired accuracy, i.e., $K \sim O[\log (1/\epsilon)]$. In total, the time complexity is polynomial to the system size and quasi-polynomial to the accuracy. Moreover, the result is greatly affected by both the initial state and the ratios of ground state energy and other eigenenergies. It is shown in Eq.~\ref{estimated_energy} that the result accuracy is proportional to the overlap between the initial state and the ground state. While fixing the result accuracy, the step $K$ should be a logarithm of inverse state overlap, i.e., $K \sim O[\log(1/|b_{0}|^{2})]$. If a random initial state is adopted, the overlap of the initial state and ground state may exponentially decay as system size increases. It indicates the iteration step $K$ should be linear with the system size. The time complexity then is proportional to $O(\epsilon^{-N})$, which has an exponential scaling to the system size under a constant $\epsilon$. This problem exists in operator-performing-based algorithms and QPE-related algorithms. A well-prepared initial state with prior knowledge of the system or assisted by VQE may help avoid this problem; the overlap being a polynomial of the system size can ensure a polynomial time complexity. As the energy ratios are dependent on the shift-energy applied, a roughly estimated ground energy from the prior knowledge also can increase the performance for solving gapped models.

Compared to the existing classical algorithms to solve eigenvalue problems whose time complexity is polynomial to the matrix size~\cite{householder1958unitary, mises1929praktische}, this quantum algorithm has an exponential speed-up over the system size for solving the ground state of a gapped system under the local Hamiltonian assumption within a fixed accuracy, if a well-prepared initial state is offered. For instance, the time complexity and space complexity are in the order of $O((2^{50})^{2}) \approx O(10^{30})$ for classical algorithms to solve $50$-qubit problems, which exceeds the power of best-known classical computers~\cite{RevModPhys.86.153, friedenauer2008simulating}. But it only requires $50$ qubits, a single qumode, and a time complexity that is polynomial to system size and a quasi-polynomial of accuracy for this quantum algorithm with a well-prepared initial state.

\section{Numerical results}
\label{Sec Numerical result}

\begin{figure}
    \centering
    \subfigure[]{
        \begin{minipage}[b]{0.45\textwidth}
            \includegraphics[width=1\textwidth]{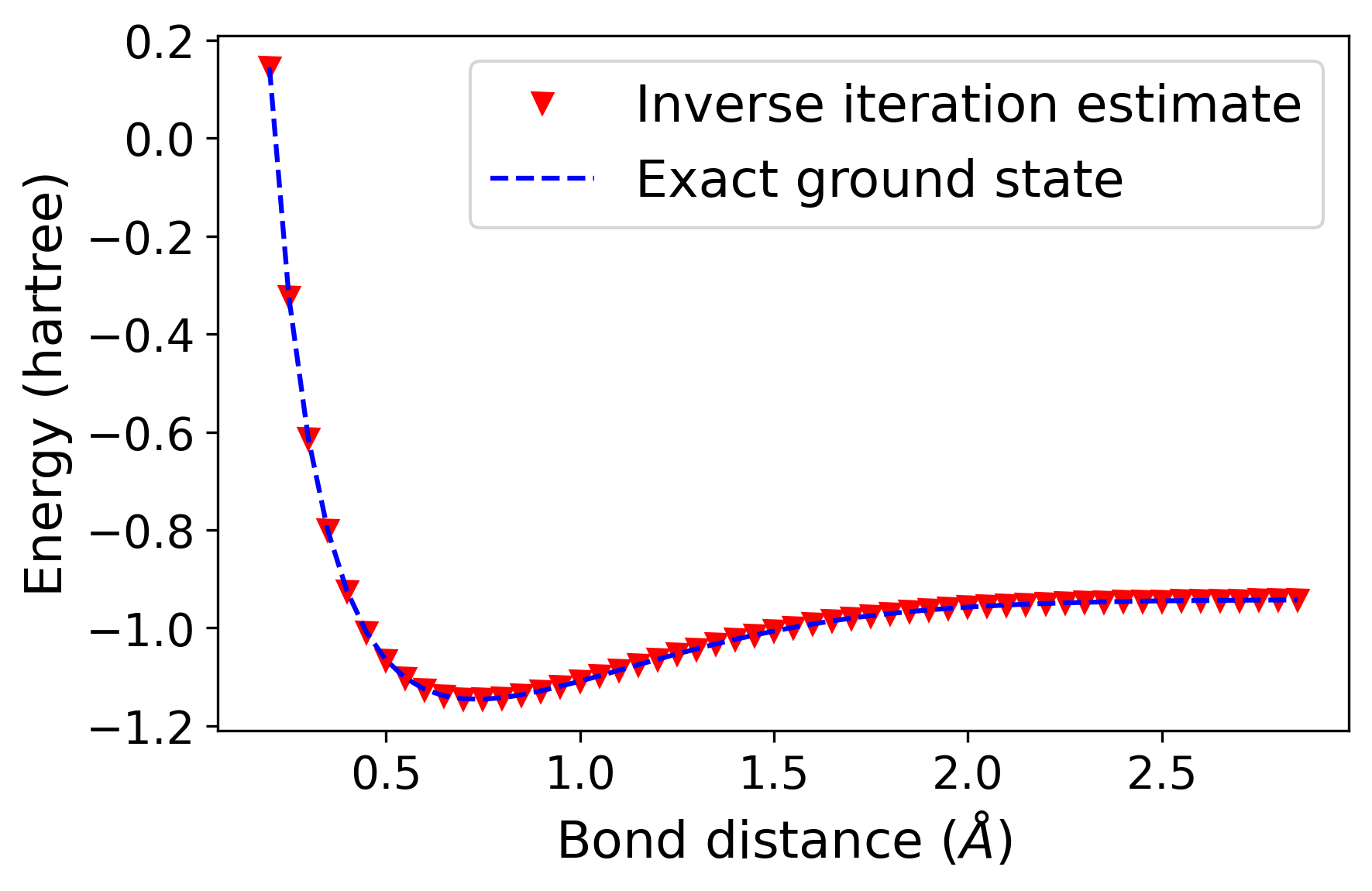}
        \end{minipage}
        \label{H2 energy}
    }
    \subfigure[]{
        \begin{minipage}[b]{0.45\textwidth}
            \includegraphics[width=1\textwidth]{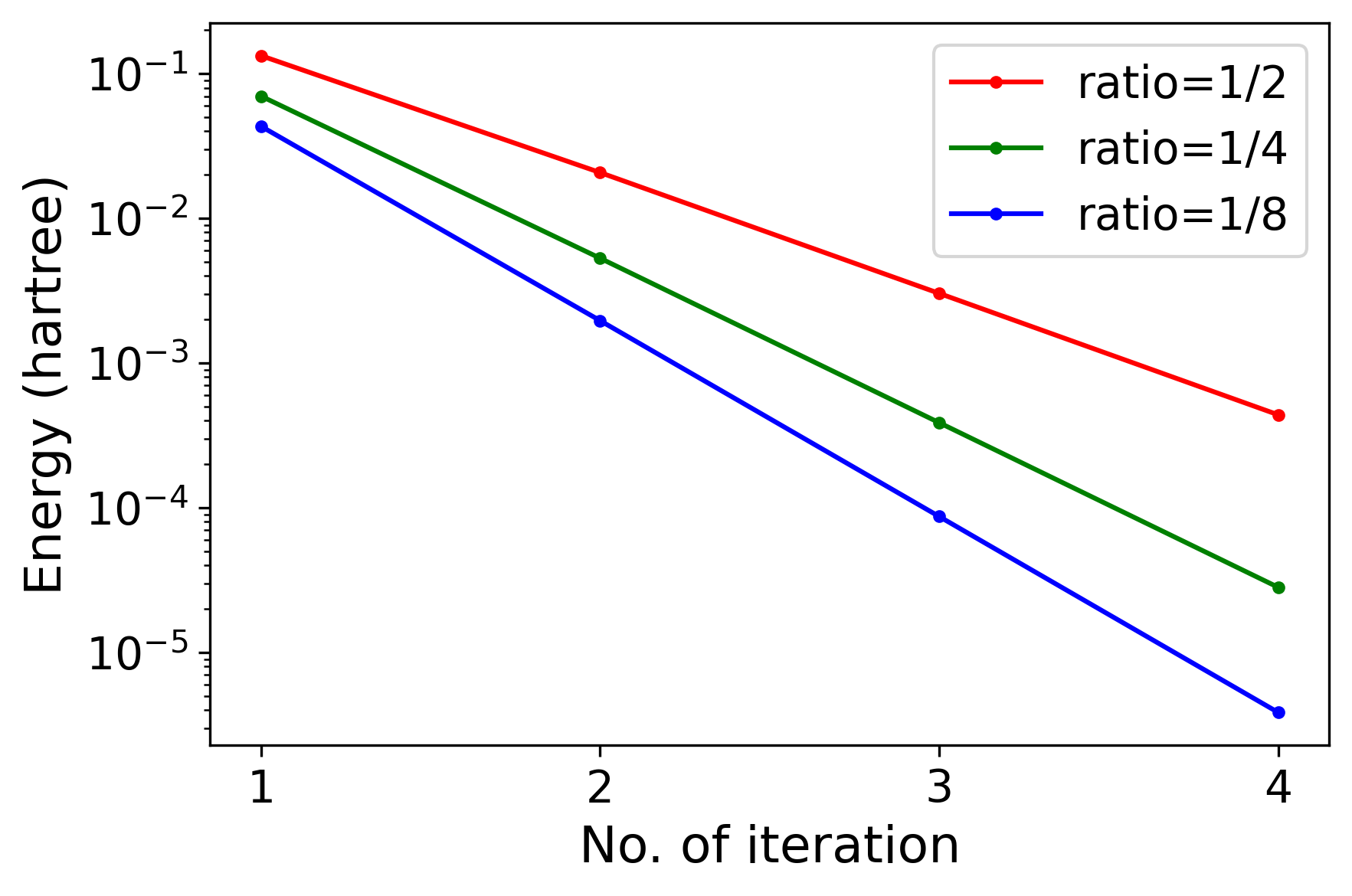}
        \end{minipage}
        \label{H2 error}
    }

    \caption{(Color online) Solving ground state energy of $H_{2}$. (a) Energy of $H_{2}$ at different bond distance. The blue dashed line is exact energy solved by diagonalizing the hydrogen molecular Hamiltonian. The red marker is solved by our quantum algorithm. (b) The energy difference between exact energy and estimated one $|E_{estimation} - E_{exact}|$ after every iteration for three chosen ratio of ground energy and first excited energy. The red line, green line and blue line correspond to the ratio being equal to $1/2$, $1/4$ and $1/8$.}
    \label{H2 simulation}
\end{figure}

\begin{figure}
    \centering
    \subfigure[]{
        \begin{minipage}[b]{0.45\textwidth}
            \label{H2_different_s_error}
            \includegraphics[width=1\textwidth]{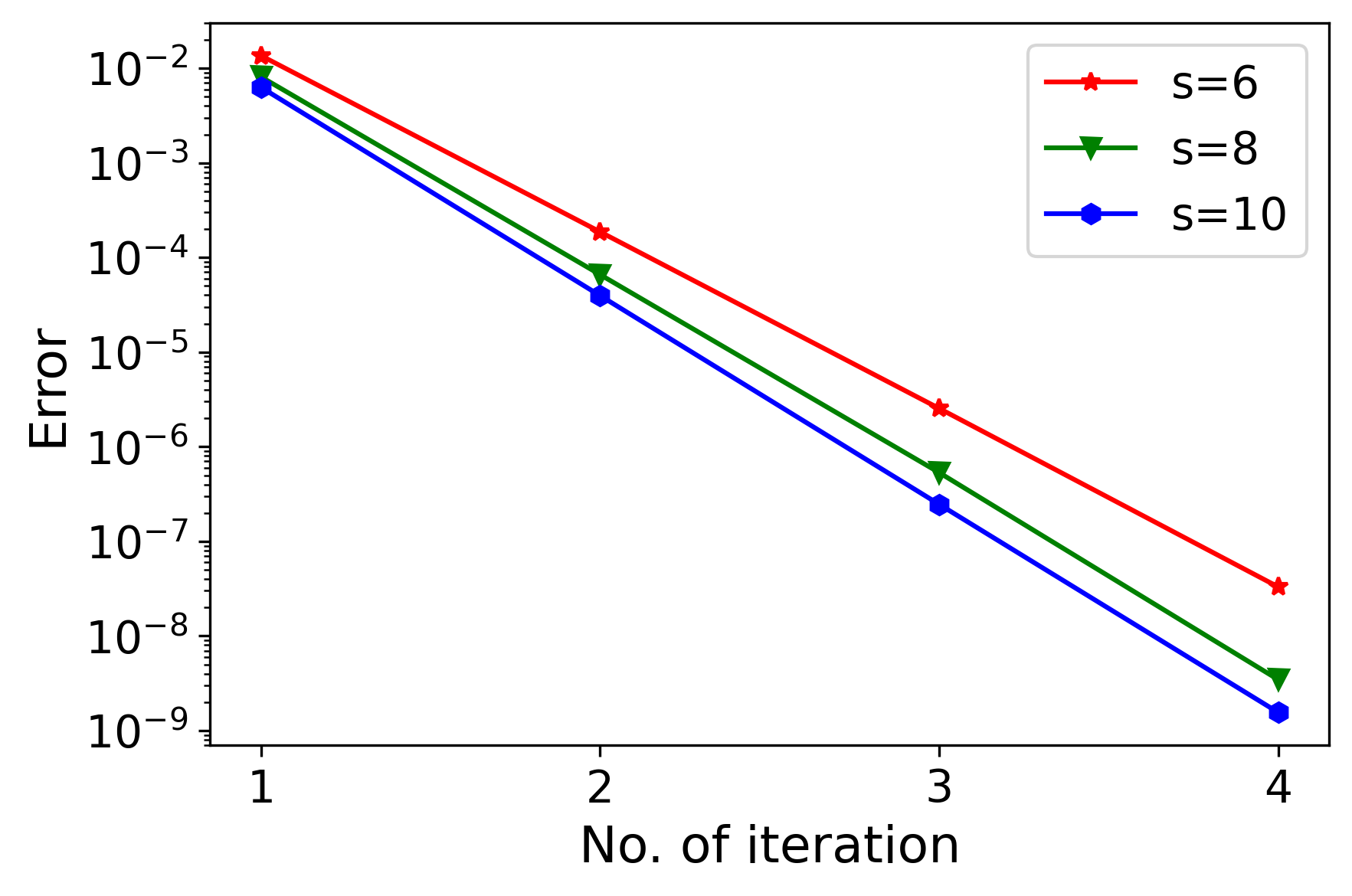}
        \end{minipage}
    }
    \subfigure[]{
        \label{H2_different_s_projection_rate)}
        \begin{minipage}[b]{0.45\textwidth}
            \includegraphics[width=1\textwidth]{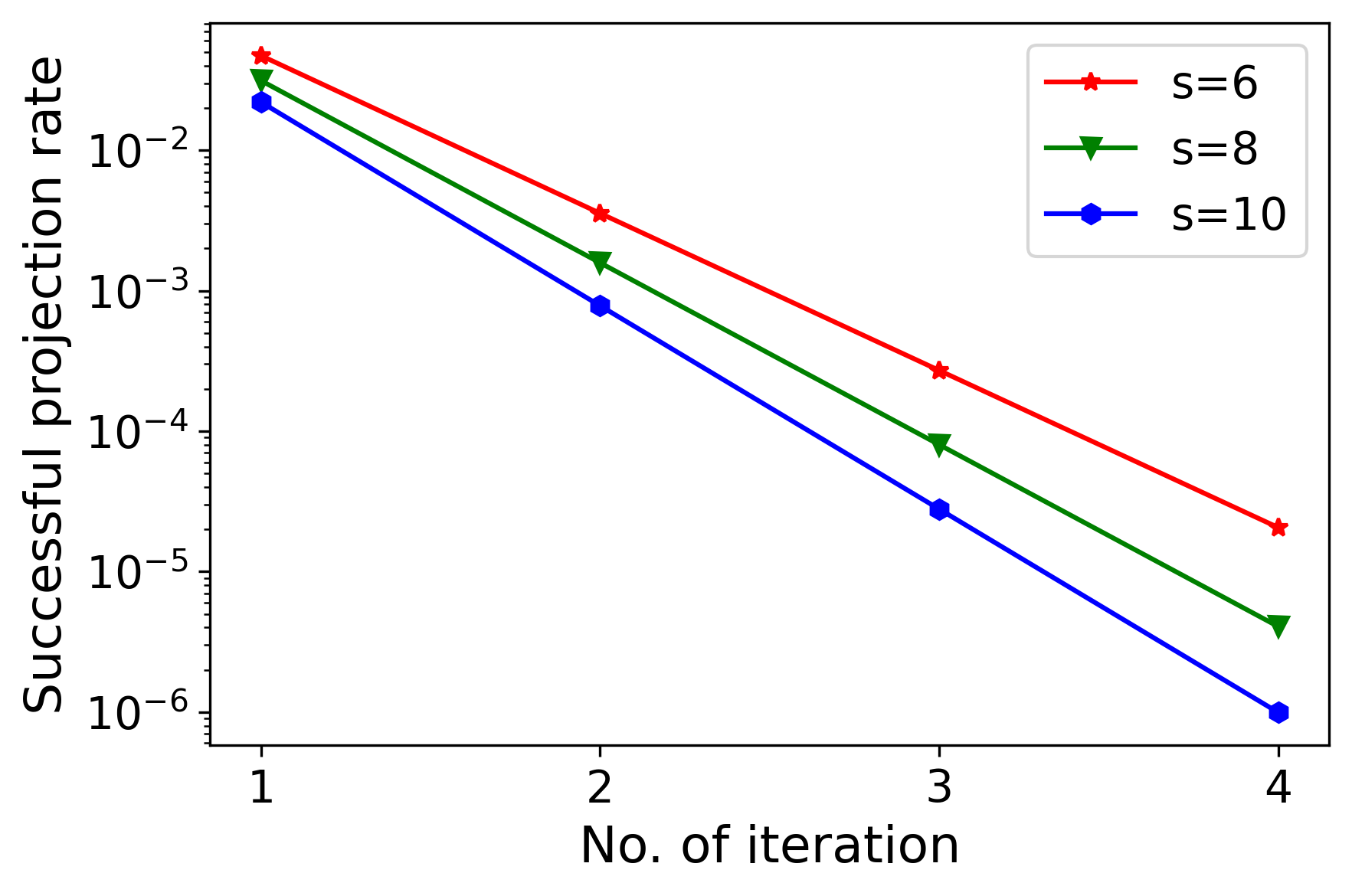}
        \end{minipage}
    }
    \caption{Solving ground state energy of $H_{2}$ for several chosen squeezing factor. Red, green, and blue lines correspond to $s = 6, 8$, and $10$ for both figures. (a) Error of estimated ground state energy after each iteration. (b) Cumulative successful projection rate after each iteration.}
    \label{H2_qumode_parameter}
\end{figure}

\begin{figure}
    \centering
    \subfigure[]{
        \begin{minipage}[b]{0.45\textwidth}
            \label{H2_noisy_loss}
            \includegraphics[width=1\textwidth]{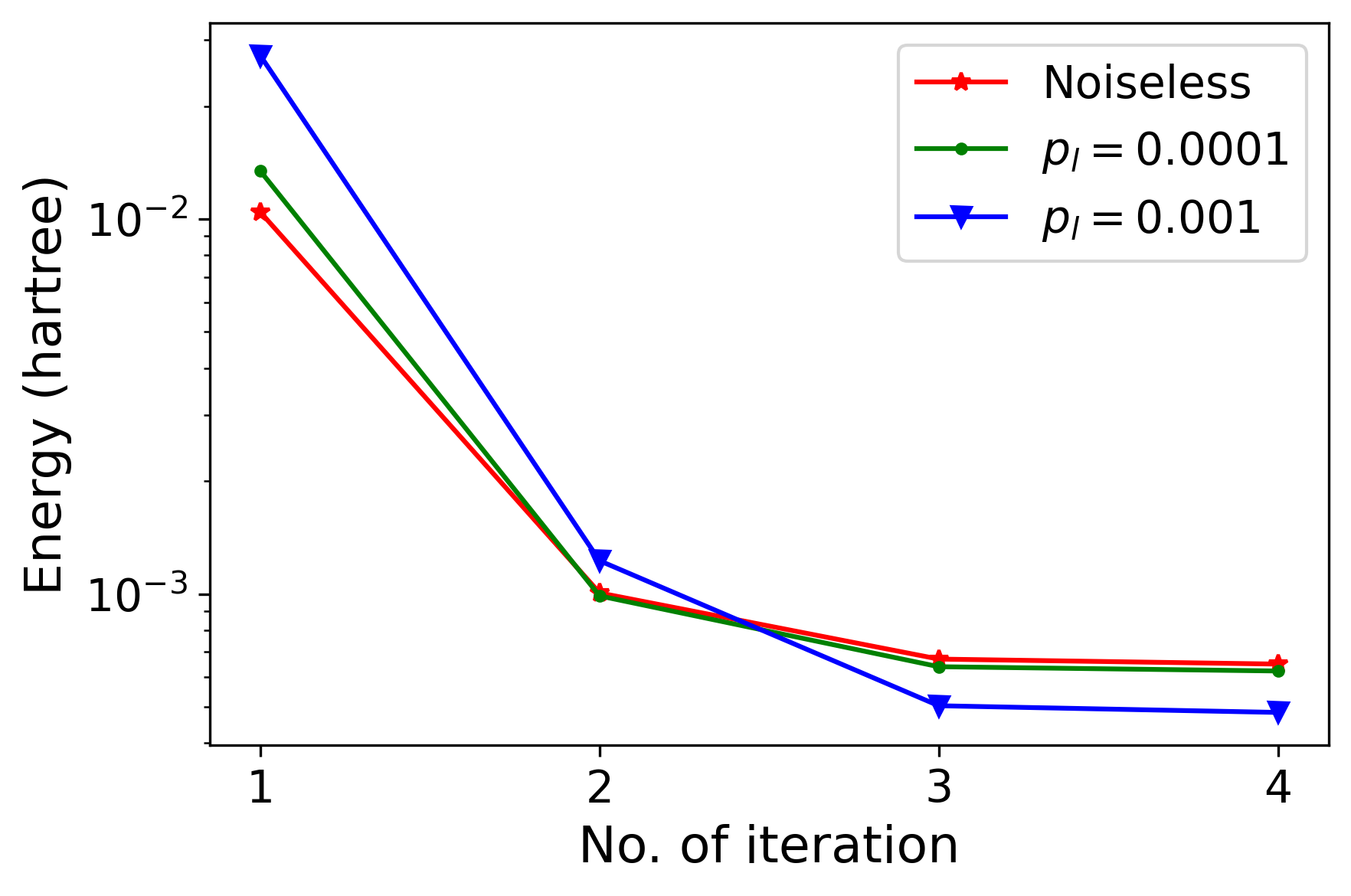}
        \end{minipage}
    }
    \subfigure[]{
        \label{H2_noisy_dep}
        \begin{minipage}[b]{0.45\textwidth}
            \includegraphics[width=1\textwidth]{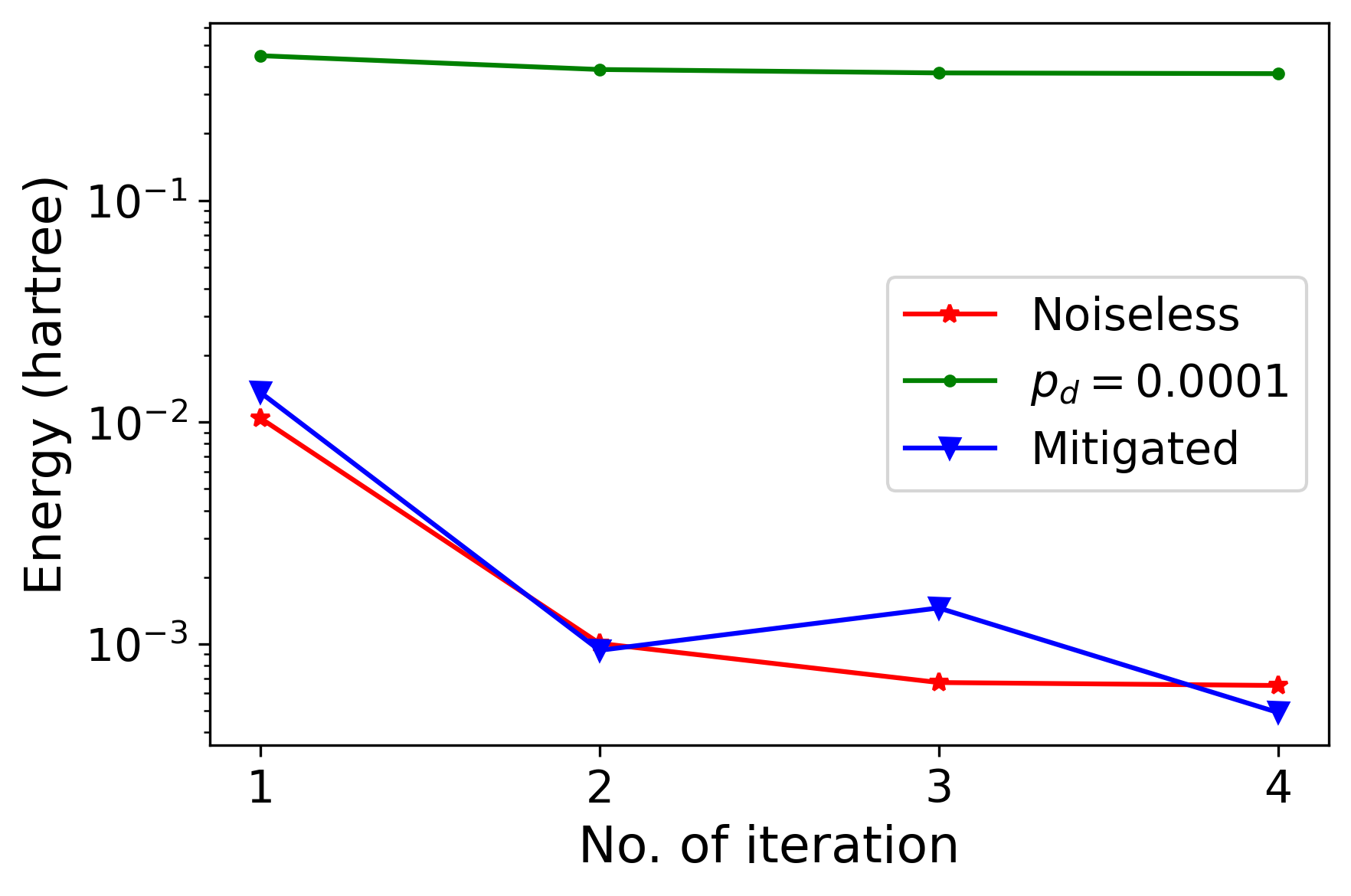}
        \end{minipage}
    }
    \caption{Solving ground state energy of $H_{2}$ for a single bond length under a noisy environment. The first order Trotter decomposition is considered and the noise affects the state after each gate. (a) Energy error after each step under two different scale lossing bosonic noise. The red, green, and blue lines correspond to noiseless case, probability of lossing boson after each gate being 0.0001 and 0.001, respectively. (b) Energy error after each iteration under depolarization noise with a probability being 0.0001. The red, green, and blue lines correspond to noiseless case, noisy case, and mitigated case, respectively.} 
    \label{H2_noisy}
\end{figure}

In this section, we simulate the QuIPI for several models, including molecular hydrogen, quantum Ising model, and Kitaev ring. They are the standard models of quantum chemistry, many-body spin system, and many-body fermionic model, respectively. This simulation is based on  \textit{QuTip}~\cite{johansson2012qutip}. The numerical results show: 1. converge rate increases as the ratio of first excited energy and ground energy increases, and iteration step can be very small for an appropriately chosen shift of energy; 2. a higher squeezing factor leads to a higher accuracy but a lower successful projection rate, namely more repetition times required; 3. QuIPI shows some robustness to noise on qumode, and error mitigation can be useful for noises on qubits; 4. the energy error is proportional to the negative quadratic step of Suzuki-Trotter decomposition $n^{-2}$, fitting the theoretical expectation; 5. phase transition can be well demonstrated by QuIPI.

\textbf{Hydrogen molecular.} Following Ref.~\cite{kandala2017hardware} that transforms the $H_{2}$ Hamiltonian to a spin system by binary tree transformation, we use two qubits to simulate the molecular hydrogen. The effective Hamiltonian presented by Pauli operators is constructed at different bond distances with parameters from Ref.~\cite{o2016scalable}, which can be expressed as, 
\begin{eqnarray}\label{ham:H2}
\hat{H}(\lambda)&=&c_0(\lambda)\hat{\mathcal{I}}+c_1(\lambda)\hat{\sigma}_1^z+c_2(\lambda)\hat{\sigma}_2^z +c_3(\lambda)\hat{\sigma}_1^z\hat{\sigma}_2^z \nonumber \\
&+&c_4(\lambda)\hat{\sigma}_1^x\hat{\sigma}_2^x+c_5(\lambda)\hat{\sigma}_1^y\hat{\sigma}_2^y.
\end{eqnarray}
We set the squeezing factor $s=10$, cut of the resource state equal to $20$, iteration step $K=3$ to estimate ground state energy. The initial state is prepared as $\ket{b}^{(0)} = \frac{1}{\sqrt{2}} (\ket{01} - \ket{10})$. The bond dissociation is shown in Fig.~\ref{H2 energy}. The estimated ground energy is perfectly fitted to the exact one.

Then the relation between converges rate and ratio of first excited energy and ground energy is investigated. Fixing bound distance at 0.75$\AA$, the ground energy and first excited energy equal to -1.15$E_{h}$ and 0.45$E_{h}$. The ratio of first excited energy and ground energy equals 2, 4, and 8 by applying shift of energy, 2.74$E_{h}$, 1.68$E_{h}$, and 1.37$E_{h}$. Fig.~\ref{H2 error} shows the error $|E_{exact}-E_{estimation}|$ after every iteration for all the three cases. The result shows that a higher ratio corresponds to the faster converge rate. Moreover, even for a bad case that ratio $ = 0.25$, the estimated energy reaches the chemical accuracy after only $3$ steps, showing $K$ may be small in practice.

Squeezed state plays an important role in the QuIPI. We discuss the effect of the squeezing factor by numerically simulating the process that is to solve the approximate ground state energy of $H_{2}$ with different squeezing factors. As shown in Fig.~\ref{H2_qumode_parameter}, increasing the squeeze factor can raise the accuracy but also leads to more samples required corresponding to a lower successful projection rate.

We also demonstrate the above process that solves the approximate ground state energy of $H_{2}$ for a fixed bond distance under a noisy environment. Here we decompose the unitary operator into several gates by the first order Trotter decomposition and consider the noise affecting the quantum state after every quantum gate. We firstly discuss the lossy bosonic noise on the qumode~\cite{PhysRevA.56.1114} after each gate, whose Kraus operator-sum representation is
\begin{equation}
    \L (\rho) = \sum_{k = 0}^{\infty} E_{k} \rho E_{k}^{\dagger},
\end{equation}
where $E_{k} = (\frac{1}{\sqrt{k!}}) p_{l}^{k/2} (1 - p_{l})^{\hat{a}^{\dagger} \hat{a}/2} \hat{a}^{k}$ with losing boson probability $p_{l}$ and creation (destroy) operator $\hat{a}^{\dagger}$ ($\hat{a}$). Fig.~\ref{H2_noisy_loss} shows this algorithm is robust to the noise on the ancillary qumode. It solves the ground state energy within the chemical accuracy even under a high-scale noise $p_{l} = 0.001$. We also simulate the process under a depolarization noise on the qubits. In Fig~\ref{H2_noisy_dep}, noise greatly affects the result. Luckily, the zero-noise limit extrapolation method can efficiently mitigate the noise in the short-depth quantum circuits without any ancilla~\cite{PhysRevLett.119.180509}. It is also shown in Fig.~\ref{H2_noisy_dep} that the energy estimated by QuIPI under a high-level noise with error mitigation can reach the chemical accuracy.

\textbf{Quantum Ising model with transverse field.} In the one-dimensional quantum Ising model, the interaction of sites is presented as a tensor product of Pauli-Z operators on the two interacted neighboring sites, and the transverse field is expressed as a Pauli-X operator performing on the single site~\cite{pfeuty1970one}. The Hamiltonian is
\begin{equation}
    H = \sum_{i=1}^{N} a_{i} \hat{\sigma}_{i}^{x} + \sum_{i=1}^{N} \sum_{j=1}^{i-1} J_{ij} \hat{\sigma}_{i}^{z} \hat{\sigma}_{j}^{z},
\end{equation}
where $N$ is the number of sites (qubits), $J_{ij}$ is interaction strength between sites $i$ and $j$, and $a_i$ is the external transverse field strength on site $i$.

\begin{figure}
    \centering

    \subfigure[]{
        \begin{minipage}[b]{0.45\textwidth}
            \includegraphics[width=1\textwidth]{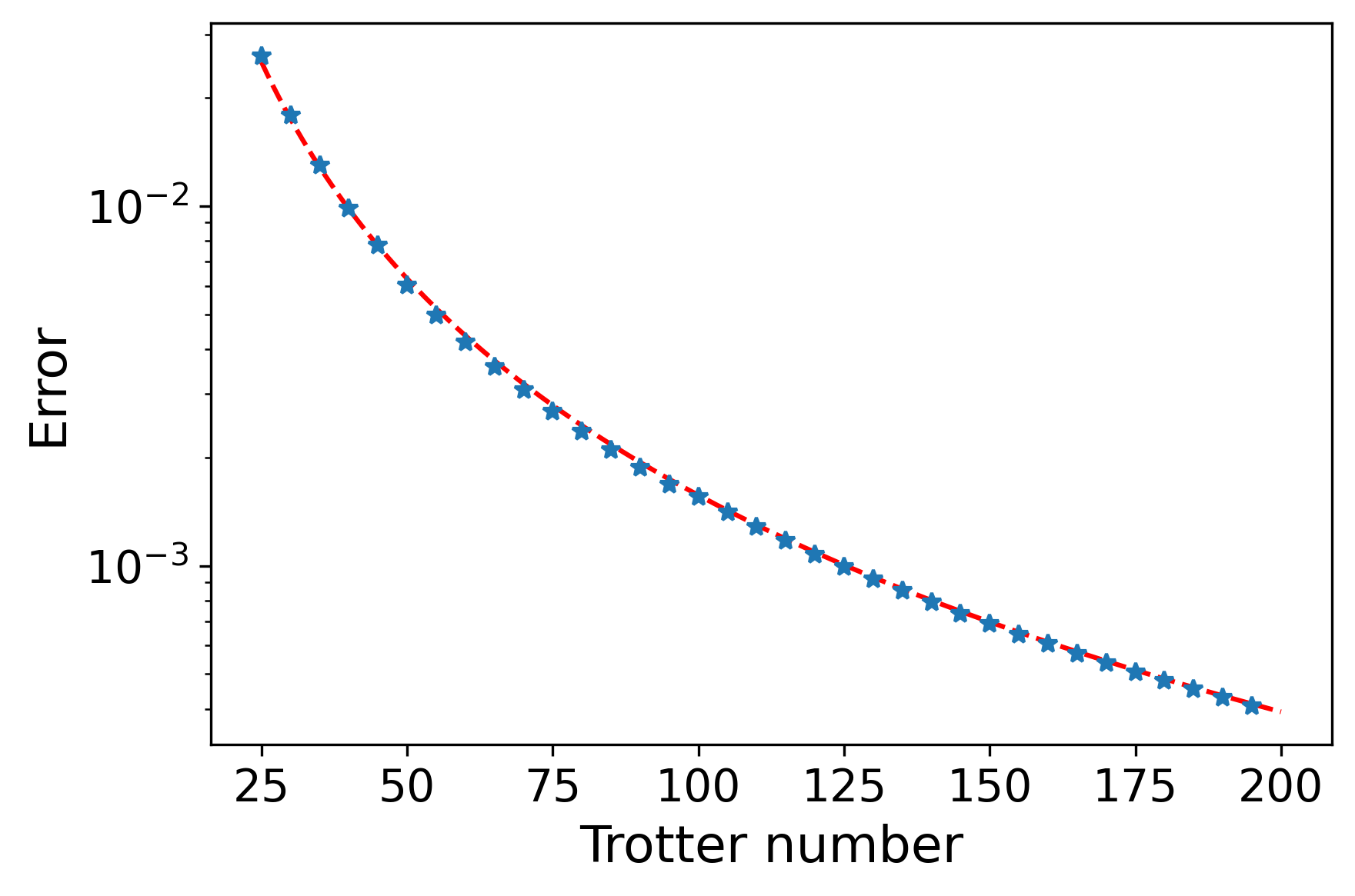}
        \end{minipage}
        \label{Ising one fidelity}
    }
    \subfigure[]{
        \begin{minipage}[b]{0.45\textwidth}
            \includegraphics[width=1\textwidth]{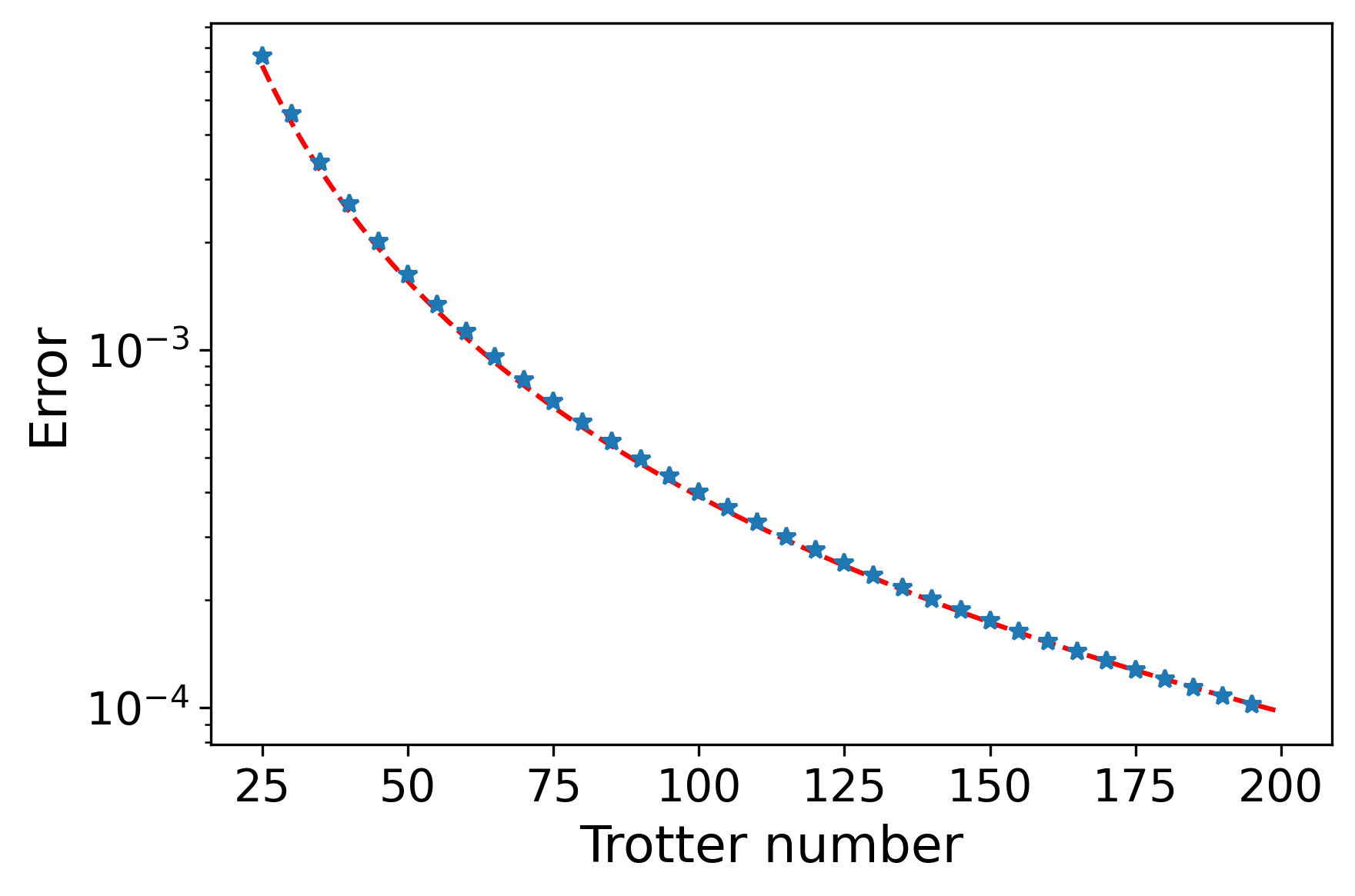}
        \end{minipage}
        \label{Ising one fidelity}
    }
    \caption{ Solving ground state energy of quantum Ising model with transverse field. 
    The blue star is error of the energy estimated by our quantum algorithm $|E_{estimation} - E_{exact}|$ and the red dashed line is the fitting curve that is proportional to $1/n^{2}$ with Trotter number $n$. (a) All the parameters $a_{i}$ and $J_{ij}$ are one. (b) Parameters are randomly sampled from 0 to 1 with an uniform distribution.}
    \label{Ising simulation}
\end{figure}

We use three qubits to solve the ground state energy of the quantum Ising model with a transverse field. Two cases are considered: 1. all the parameters equal to one; 2. all the parameters are randomly chosen from a uniform distribution $[0, 1]$. The initial state is prepared by applying a Hadamard gate and a $\hat{\sigma}^z$ gate on each qubit, i.e., $\ket{b}^{(0)} = (\hat{\sigma}^{z})^{\otimes n} H^{\otimes n} \ket{0}^{\otimes n}$. We analyze the relation between the energy error and the number of Suzuki-Trotter decomposition. The result is shown in Fig.~\ref{Ising simulation}, in which the blue star is energy errors at different Trotter number $n$ and red dashed line is the fitting function that is proportional to $1/n^{2}$, meeting theoretical expectation mentioned in Sec.~\ref{Time complexity analysis}.

It should be pointed out that solving a general quantum Ising model of large size with random coupling can be NP-hard, and it is not expected that QuIPI can solve it in general either.

\textbf{The Kitaev ring.} The Kitaev ring is an one-dimensional fermion system that can be used to demonstrate quantum phase transition~\cite{kitaev2001unpaired}. The Hamiltonian is 
\begin{equation}
    \hat{H} = -J \sum_{i=1}^{N} (\hat{c}_{i}^{\dagger} \hat{c}_{i+1} + \hat{c}_{i}^{\dagger}\hat{c}_{i+1}^{\dagger} + h.c.) - \mu \sum_{i=1}^{N} \hat{c}_{i}^{\dagger} \hat{c}_{i}.
\end{equation}
Mapped to spin form by Jordan-Wigner transformation~\cite{jordan1993paulische}, it is presented as
\begin{equation}
    \hat{H} = - h \sum_{i=1}^{N} \hat{\sigma}_{i}^{z} - J \sum_{i=1}^{N-1} \hat{\sigma}_{i}^{x} \hat{\sigma}_{i+1}^{x} - J \hat{\sigma}_{1}^{y} (\Pi_{i=2}^{N-1} \hat{\sigma}_{i}^{z}) \hat{\sigma}_{N}^{y}.
\end{equation}

\begin{figure}
    \centering
    \subfigure[]{
        \begin{minipage}[b]{0.45\textwidth}
            \includegraphics[width=1\textwidth]{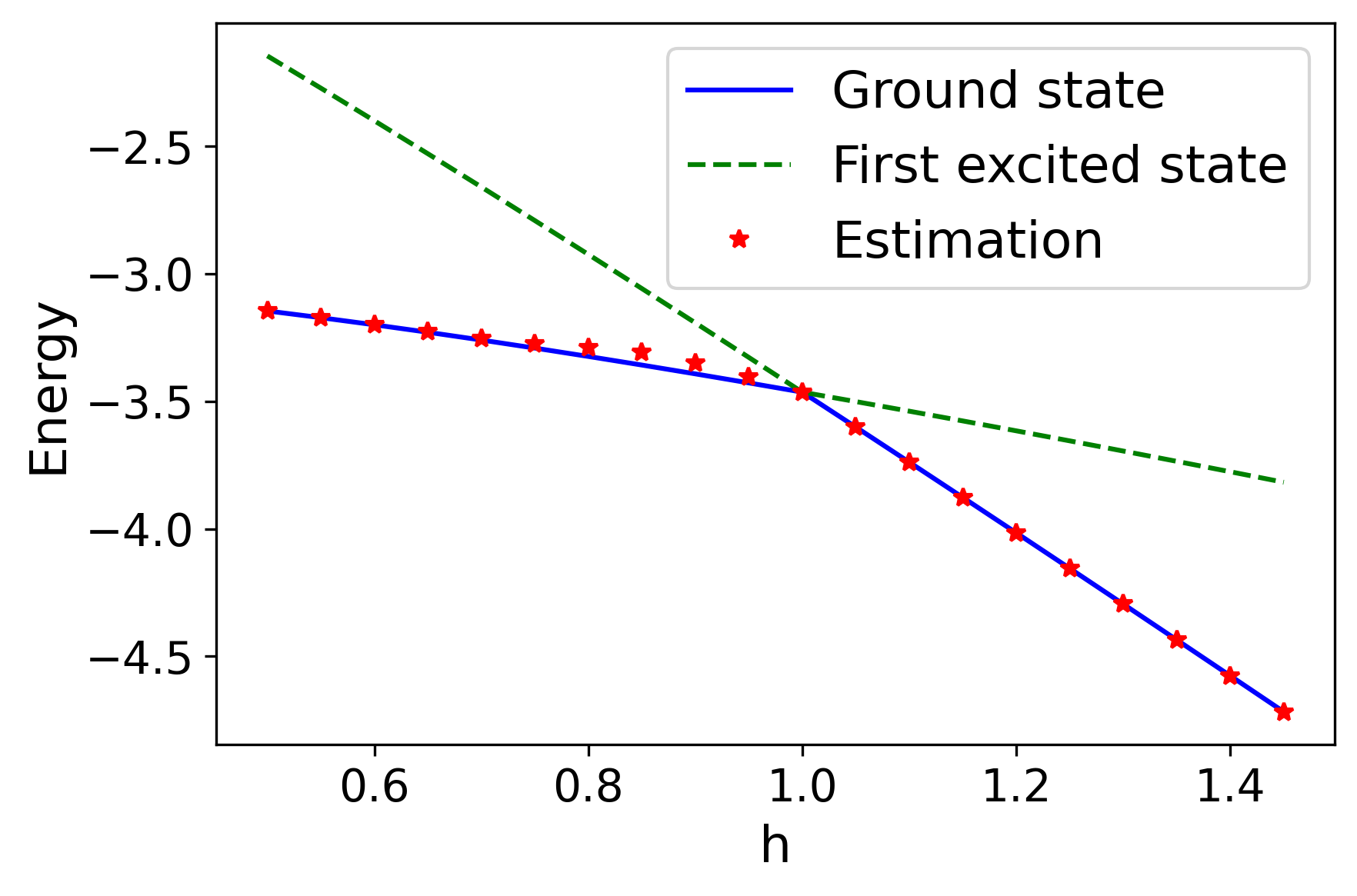}
        \end{minipage}
        \label{Kitaev fidelity}
    }
    \caption{(Color online)~Solving ground state energy of Kitaev model. The blue line and green line are the exact ground state energy and first excited state energy, and red star is the energy estimated by our quantum algorithm. At $h = 1$, quantum phase transition happens.} 
    \label{Kitaev simulation}
\end{figure}

This model with $J = 1$ and changing $h$ is simulated by three qubits. We prepare the initial state as $\ket{b}^{(0)} = H^{\otimes n} \ket{0}^{\otimes n}$ ($\ket{b}^{(0)} = \ket{0}^{\otimes n}$ ) before (after) quantum phase transition. Fig.~\ref{Kitaev simulation} shows the exact ground energy, first excited energy, and the result energy solved by our quantum algorithm. At $h = 1$, the quantum phase transition is well illustrated.

\section{Discussion and conclusion}
\label{sec Discussion and conclusion}

So far, we have focused on solving the ground eigenstate and ground state energy. However, the QuIPI and hybrid quantum-classical strategy can solve not only the ground state energy but also the whole energy spectrum. As we mentioned, the resulting state of the iterative power iteration method corresponds to the minimum absolute eigenvalue. So, we can determine any energy level by applying a shift of energy near the target energy. In this situation, the energy ratio mentioned in the time complexity should be replaced by the ratio of the target energy and the second minimum absolute energy. In addition, as the result accuracy depends on the energy ratio, the performance of QuIPI may increase if the shift-energy applied makes the value of ground state energy small. Thus, a strategy of iteratively updating the shift-energy may be incorporated into the QuIPI.

Our algorithm takes advantage of continuous-variable resources. To show it, we make a comparison between the QuIPI and the hybrid quantum-classical IPI that does not use the continuous-variable qumode.  The hybrid one realizes inverse Hamiltonian by performing a series of unitary operators with different evolution time, i.e., $\hat{H}^{-1} \approx \sum_{j=0}^{M_j-1} e^{-i\hat{H} j \Delta p} \Delta p$, where $\Delta p$ is the discrete interval of the summation, and $M_{j}$ is the up limit of the summation (see~\ref{Sec Hybrid quantum-classical method} for details as well as numerical results for hydrogen molecular). This form implies that the hybrid quantum-classical IPI demands the long-time evolution of $\hat{H}$, namely, $t$ in $e^{-i\hat{H}t}$ should be large enough. In comparison, the QuIPI refers to a qumode to encode weights of different unitaries into the resource state. As the resource state naturally has distribution for large quadrature $\hat{p}$, the long-time evolution of the Hamiltonian is intrinsically realized with a coupling between the system and the qumode in terms of $e^{-i\hat{H}\hat{p}}$. This comparison shows that the continuous-variable resource used in QuIPI can reduce the demanded coherence evolution time.

We briefly discuss the physical implementation of the quantum algorithm since pursuing the advantage of continuous-variable in QuIPI relies on a hybrid-variable quantum platform for physical implementation. We note that the mainstream platforms of quantum computers based on qubits, such as trapped ions and superconducting circuits, often have continuous variables that couple with qubits. The physical implementation is very similar to Ref.~\cite{zhang2020continuous} utilizing an auxiliary qumode, which is feasible for current quantum platforms, in the sense that all necessary components are readily implementable, including preparing the resource state $\ket{R,s}$, implementing $e^{-i\hat{H}\hat{p}}$ and projection.

While the universal fault-tolerant quantum computers have not been developed yet, we still face the limitation of coherence time on the near-term quantum processors. The existence of decoherence restricts the circuit depth, thus limiting the accuracy of the final result solved by this algorithm. We briefly discuss the feasibility of the physical implementation of this algorithm on near-term quantum computers. From the state-of-art technique~\cite{campagne2020quantum}, the lifetime of a single photon at their quantum processor $T_{p}$ is approximate to $245 \mu s$, and the storage time of $\ket{n}$ state is about $T_{p}/n$. The storage time is at the order of $100 \mu s$ and $10 \mu s$ for the two-level qubit and the ancillary qumode with the highest Fock state below $100$, respectively. Then, the implementing time of CNOT gate $T_{c}$ and hybrid qubit-qumode gate $T_{h}$ are at the order of $100 ns$ and $10 ns$. From the rough estimation, we can see hundreds of hybrid qubit-qumode gates can be performed within the coherent time, which means the existing quantum processor can possibly solve quantum systems within tens of local operators and tens of Trotter number by our quantum algorithm. With the development of experimental technology, the larger quantum system can be solved with higher accuracy by our quantum algorithm with a longer coherent time in the near future.

In summary, we have proposed inverse iteration quantum eigensolvers for solving eigenstates of Hamiltonian, which utilizes a continuous variable qumode to realize inverse Hamiltonian as an integral of Hamiltonian evolution. 
We have demonstrated the efficiency and accuracy of the QuIPI for a range of quantum systems, including both quantum chemistry and quantum many-body models. We also have proposed a hybrid quantum-classical algorithm of IPI, where the integral is discretized, and unitaries are summed classically. Compared with QuIPI, the hybrid algorithm relies on a long time for Hamiltonian evolution. Lastly, we point out that the quantum algorithm developed here may also be applied for matrix-inversion-based quantum machine learning.

\section{Acknowledgements}
We thank Guo-Qing Zhang for helpful discussions. This work was supported by the CRF of Hong Kong (No. C6005-17G), the National Natural Science Foundation of China (Grants No. 91636218 and No. U1801661), the National Key Research and Development Program of China (Grant No. 2016YFA0301800), the Key Project of Science and Technology of Guangzhou (Grant No. 201804020055), and the Key-Area Research and Development Program of GuangDong Province (Grant No. 2019B030330001). We also thank support from Guangdong-Hong Kong Joint Laboratory of Quantum Matter.

\appendix
\section{Energy error result from inverse iteration}
\label{Energy error result from inverse iteration}
The initial state can be represented as a superposition state of eigenstates of $\hat{H}$
\begin{equation}
    \ket{b} = \sum_{i} b_{i} \ket{\psi_{i}}
\end{equation}
with
\begin{equation}
    \hat{H}^{-1} \ket{\psi_{i}} = E_{i}^{-1} \ket{\psi_{i}}.
\end{equation}
So, performing $k$ times $\hat{H}^{-1}$ on initial state $\ket{b}$ leads to
\begin{equation}
    \ket{\psi_{g^{\prime}}} = \hat{H}^{-k} \ket{b} = \frac{1}{\sqrt{c}} \sum_{i} b_{i} E_{i}^{-k} \ket{\psi_{i}}
\end{equation}
with normalization factor $1/\sqrt{c}$.

The expectation value of $\hat{H}$ is
\begin{equation}
    \label{estimated_energy}
    \begin{aligned}
        \braket{\hat{H}} &= \braket{\psi_{g^{\prime}}|\hat{H}|\psi_{g^{\prime}}} \\
        &= \frac{1}{c} \sum_{i,j} b_{i}^{\ast} b_{j} E_{i}^{-k} E_{j}^{-k} \braket{\psi_{i}|\hat{H}|\psi_{j}} \\
        &= \frac{1}{c} \sum_{i} |b_{i}|^{2} E_{i}^{-2k} E_{i} \\
        &= \frac{1}{c} (|b_{0}|^{2} E_{0}^{-2k+1} + |b_{1}|^{2} E_{1}^{-2k+1} + \cdots) \\
        &= \frac{1}{c} E_{0}^{-2k} [|b_{0}|^{2} E_{0} + |b_{1}|^{2} (\frac{E_{0}}{E_{1}})^{2k} E_{1} + \cdots] \\
        &= E_{0} + O(|b_{0}|^{2} \cdot (\frac{E_{0}}{E_{1}})^{2k})
    \end{aligned}
\end{equation}

\section{Demonstration of resource state preparation and its effect}
\label{Demonstration of resource state preparation}
There are several methods to prepare the resource state we need, such as using continues-variable quantum neural network~\cite{killoran2019continuous}, exploiting evolution of Jaynes-Cummings type qubit-qumode coupling ~\cite{law1996arbitrary}, and sequentially applying coherent displacement operator and phonon creation operator~\cite{dakna1999generation}. We take the last method as an example to prepare our finite squeezed state to show the feasibility of resource state preparation. And we also discuss the effect of the highest Fock state considered to the resource state preparation and performance of QuIPI.

We firstly rewrite our resource state as a superposition state of Fock states:
\begin{equation}
    \ket{R,s} \approx \sum_{n=0}^{cut} c_{n} \ket{n},
\end{equation}
where cut is the highest Fock state we consider, the weights $c_{n}$ is transformed from the resource state in momentum space $c_{n}=\sqrt{2} s^{-1/2} \pi^{-1/4} \int_{0}^{\infty} e^{-p^{2}/2s^{2}} \braket{n|p} dp$ with $\braket{n|p} = i^{n} \pi^{-1/4} \frac{1}{\sqrt{2^{n} n!}} H_{n}(p) e^{-p^{2}/2}$, where $H_{n}(p)$ is Hermite polynomials.

The target state can be achieved by
\begin{equation}
    \ket{R,s} \approx \Pi_{n=1}^{cut} \hat{D}(\alpha_{n})\hat{a}^{\dagger}\hat{D}(\alpha_{n})^{\dagger} \ket{0},
\end{equation}
where $\hat{D}(\alpha_n)$ is coherent displacement operator, $\hat{a}^{\dagger}$ is creation operator, and $\left\{ \alpha_n \right\}$ is determined by
\begin{equation}
    \sum_{n=0}^{cut} \frac{c_{n}}{\sqrt{n!}} (\alpha^{\ast})^{n}=0.
\end{equation}

\begin{figure}[H]
	\centering
	\subfigure[]{
		\begin{minipage}[b]{0.45\textwidth}
			\includegraphics[width=1\textwidth]{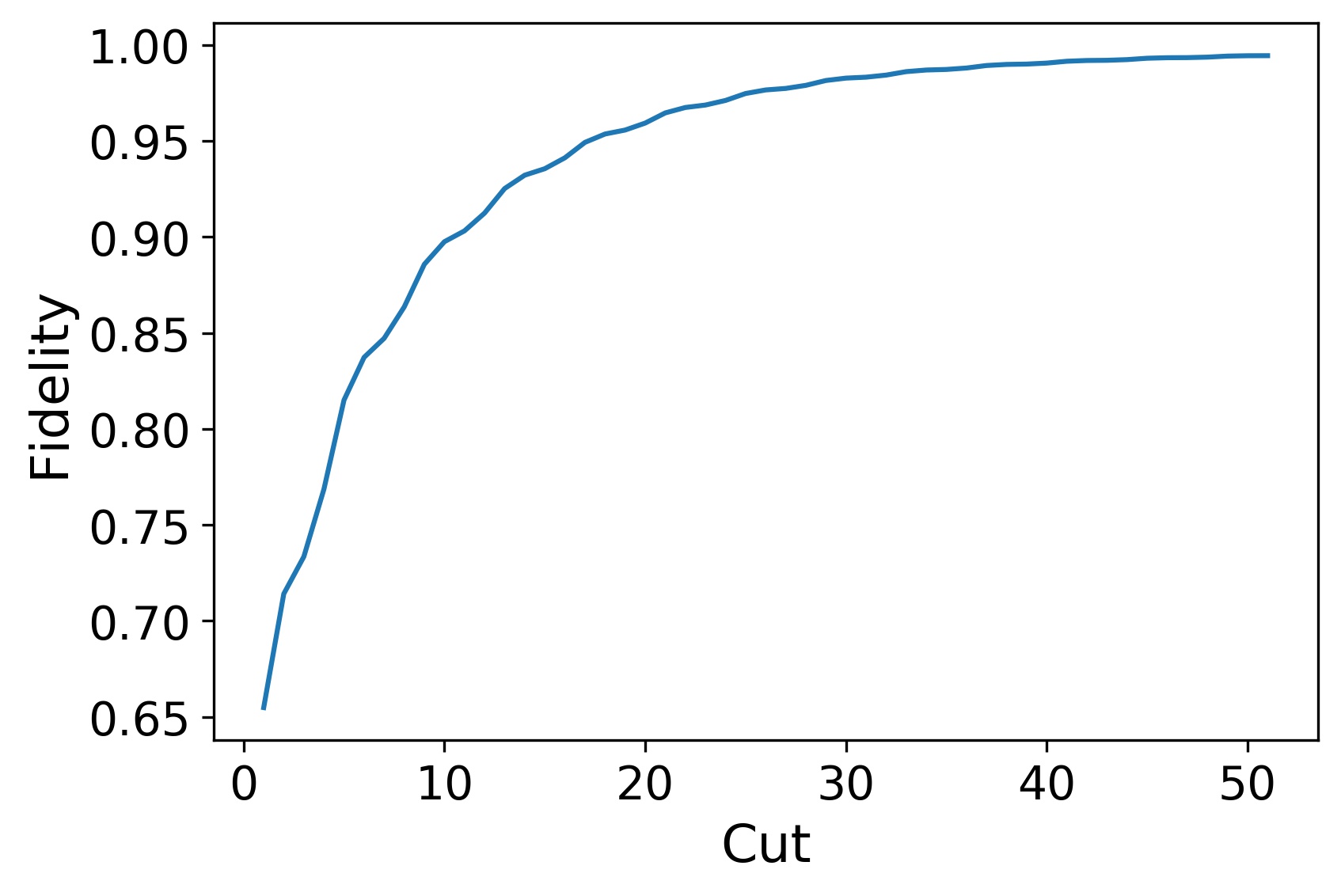}
		\end{minipage}
		\label{Fidelity vs cut}
	}
	\subfigure[]{
		\begin{minipage}[b]{0.45\textwidth}
			\includegraphics[width=1\textwidth]{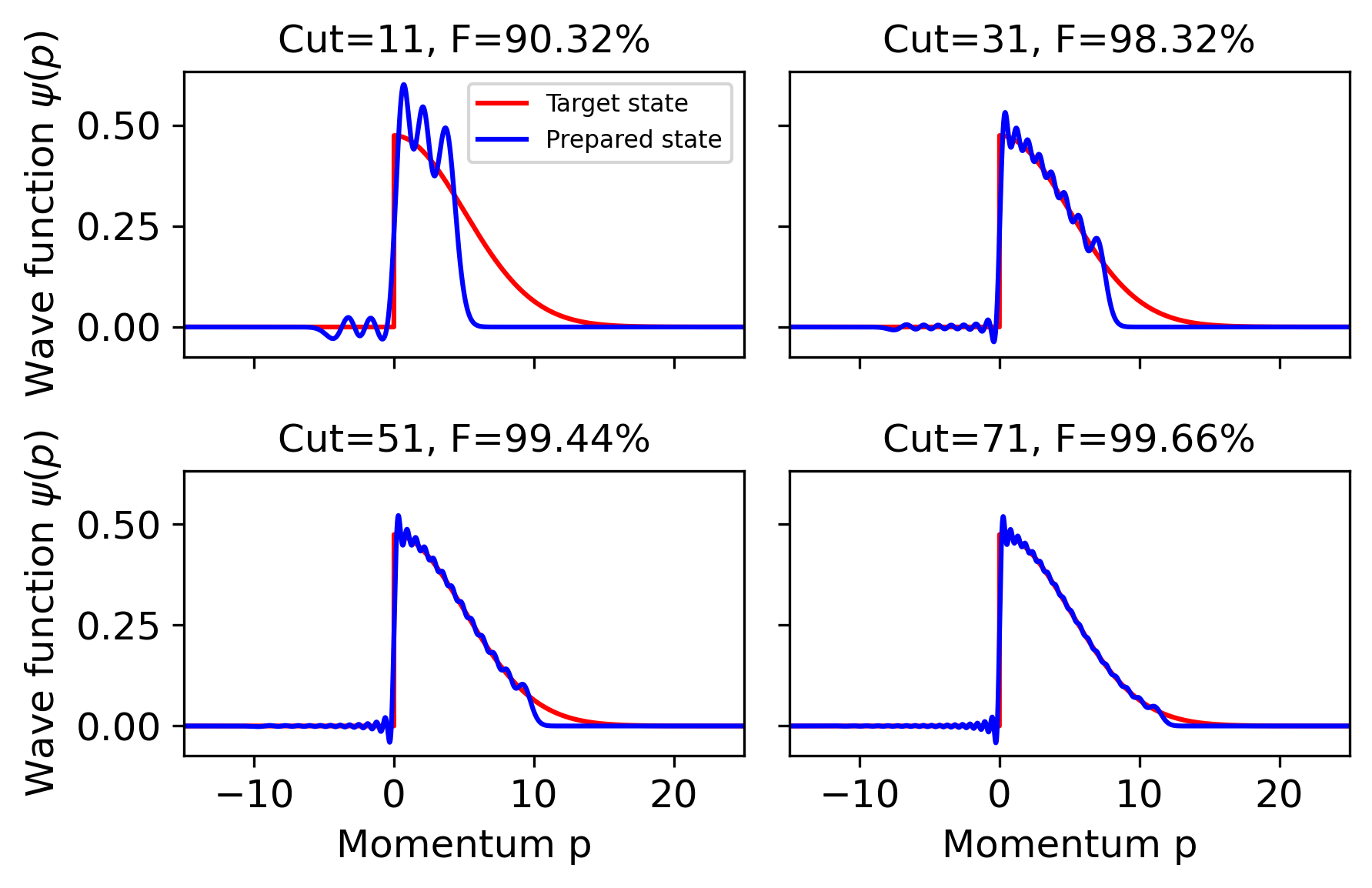}
		\end{minipage}
		\label{Momentum wavefunctions}
	}

    \subfigure[]{
		\begin{minipage}[b]{0.45\textwidth}
			\includegraphics[width=1\textwidth]{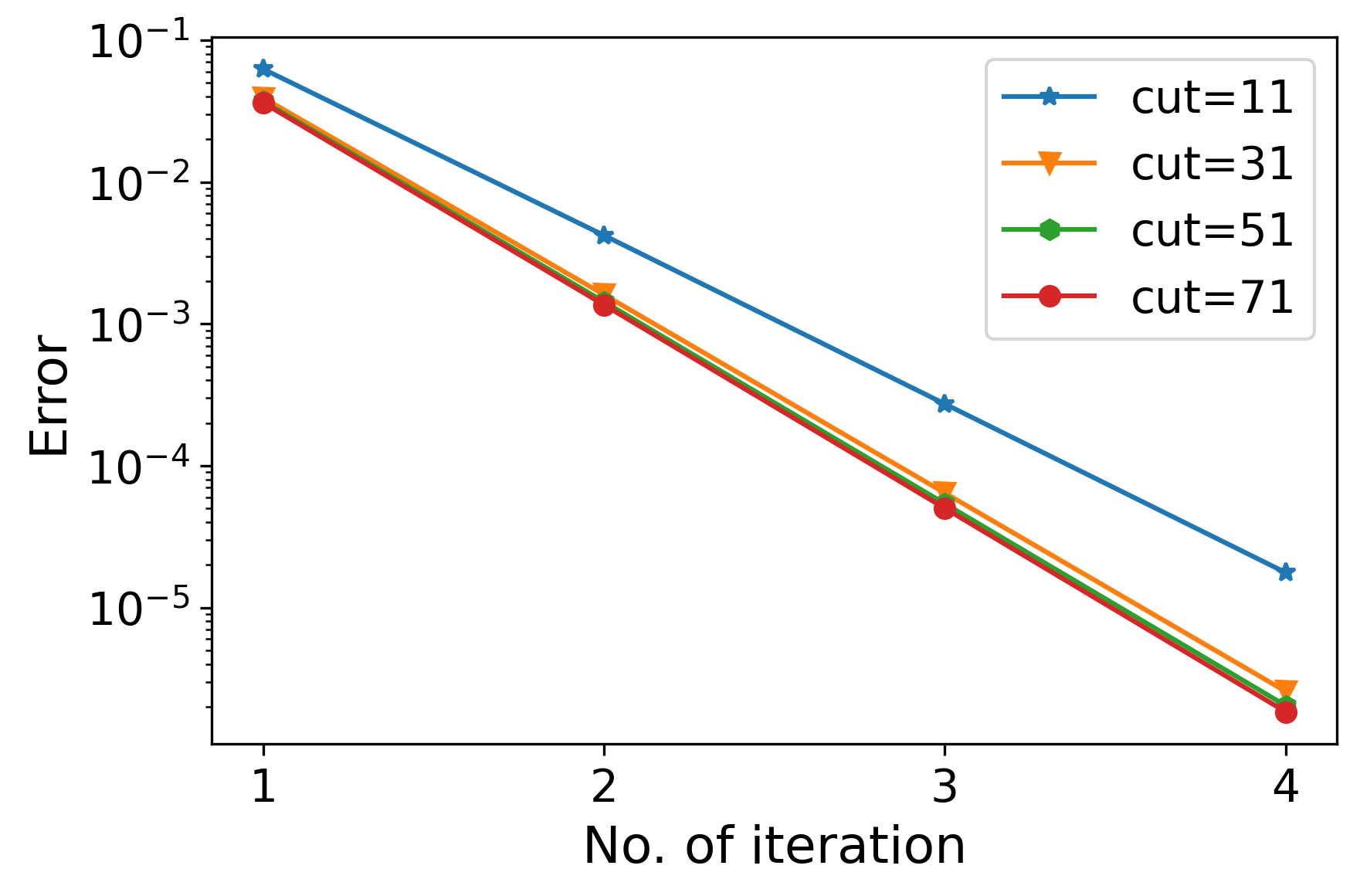}
		\end{minipage}
		\label{H2_different_cut}
	}
    \subfigure[]{
		\begin{minipage}[b]{0.45\textwidth}
			\includegraphics[width=1\textwidth]{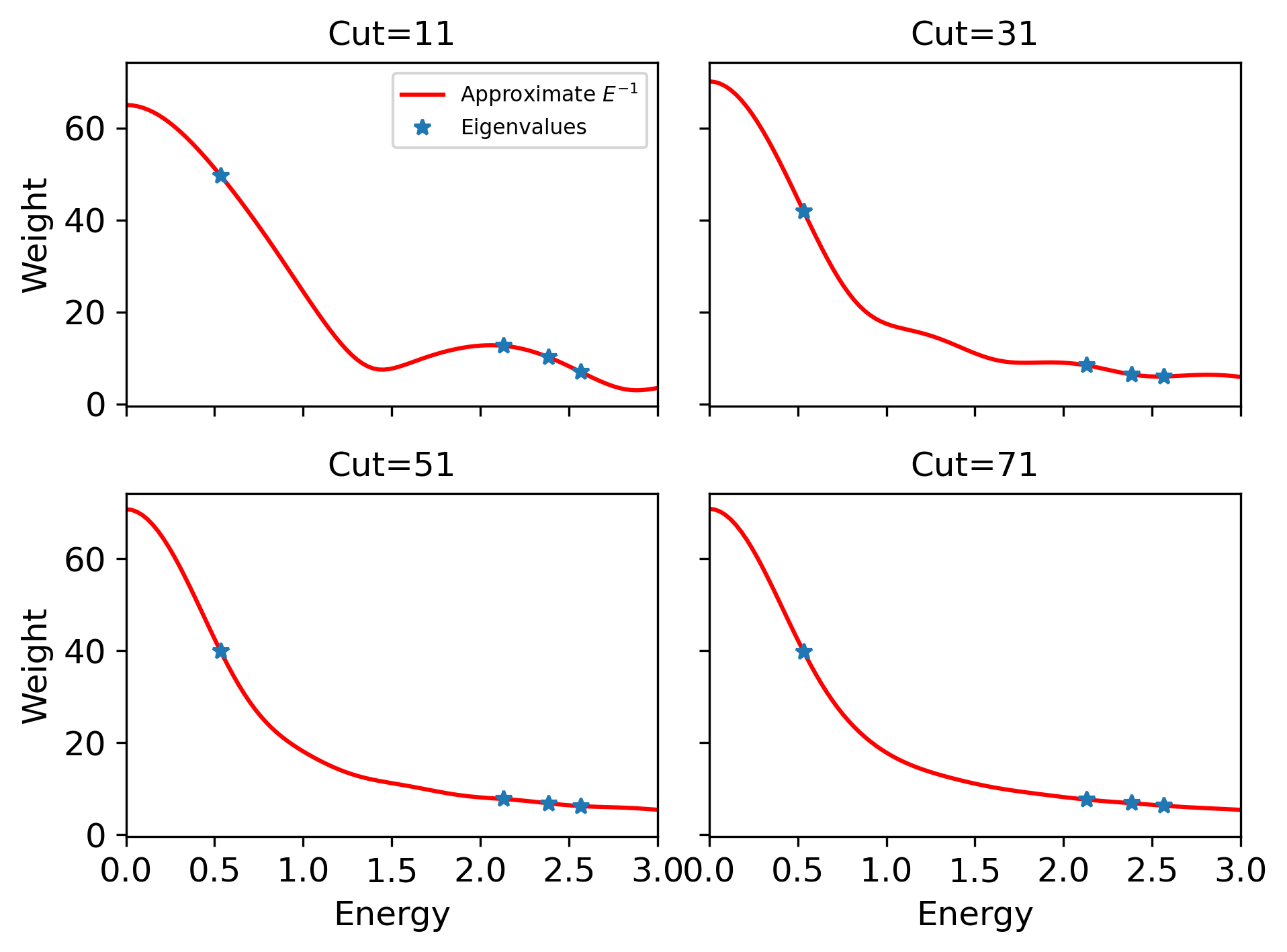}
		\end{minipage}
		\label{additional_weight_cuts}
	}
	
	\caption{Qumode resource state preparation and the effects of cut to QuIPI. (a) The fidelity between the target state and the state we prepared by the method that alternately applying the coherent displacement operator and creation operator. (b) The momentum wave function of the prepared qumode state with four chosen cut and the target momentum wave function. (c) Energy error of the ground state energy of $H_{2}$ for a fixed bond solved by QuIPI with different cut. (d) The additional weigh for single approximate inverse Hamiltonian for different cut numbers. It is a function of eigenvalue, where performing this approximate inverse Hamiltonian gives an additional weight to the corresponding eigenvector.}
	\label{Resource state}
\end{figure}

We simulate this process to prepare our qumode resource state with a squeezing factor equal to five $\ket{R,s=5}$. Fig.~\ref{Fidelity vs cut} shows increasing cut can increase the fidelity of the result qumode state. Fig.~\ref{Momentum wavefunctions} shows the momentum distribution for the four chosen cuts. Theoretically, preparing the resource state with considering higher Fock state results in higher fidelity, which leads to better performance of the QuIPI. We demonstrate the process to solve the approximate energy of $H_{2}$ for a bond distance at $0.75 \AA$, where a shift-energy $1.68 E_{h}$ is applied, the squeezing factor is $s = 5$, and four different cut numbers are chosen. It is shown in Fig.~\ref{H2_different_cut} that a minor cut has lower accuracy. It results from a bad approximation of inverse Hamiltonian. Applying a perfect inverse Hamiltonian on state results in an additional weight for each eigenstate of this Hamiltonian that has an inverse relationship to the corresponding eigenvalue. Fig.~\ref{additional_weight_cuts} shows that a higher cut makes the approximate inverse Hamiltonian more similar to the exact one. Though the larger cut number leads to better result, it is more difficult to manipulate the higher Fock state as more energy is required. Moreover, it is less robust to the noise.

\section{Finite squeezing effect}
\label{Finite squeezed result}
Squeezed states are obtained by squeezing the qumode probability distribution on the position (momentum) space and extending it on the momentum (position) space. In this process, the qumode state conforms to the uncertainty principle from beginning to end. In Sec.~\ref{Quantum version of inverse power iteration method}, the qumode states are infinitely squeezed states. The projection qumode state $\ket{q=0} = \int_{-\infty}^{\infty} \ket{p} dp$ (not normalizable) is infinitely squeezed on the position space and extended on the momentum space, so that it has a certain position $q=0$ but momentum is equally distributed from negative infinity to positive infinity. The resource state $\ket{R}=\int^{\infty}_{0} \ket{p} dp$ (not normalizable) is infinitely squeezed on position space too, but only have positive momentum. In this case, the successful projection rate is zero. Moreover, infinite squeezed states can not experimentally obtained. For these two reasons, we have to consider finite squeezed states: $\ket{q,s} = s^{-1/2} \pi^{-1/4} \int_{-\infty}^{\infty} e^{-p^2/2s^2} \ket{p} dp$ and $\ket{R,s}= \sqrt{2} s^{-1/2} \pi^{-1/4} \int_{0}^{\infty} e^{-p^2/2s^2} \ket{p} dp$.

Considering the finite squeezed state, the result state after unitary operator performed and projection is
\begin{equation}
    \begin{aligned}
        \ket{\psi^{\prime}} &= \bra{q,s}e^{-i\hat{H}\hat{p}} \ket{R,s} \ket{b} \\
        &= \sqrt{2} s^{-1} \pi^{-1/2} \sum_{n} b_{n} \int_{0}^{\infty} e^{-iE_{n}p} e^{-p^{2}/s^{2}} dp \ket{\psi_{n}} \\
        &= \frac{\sqrt{2}}{2} \sum_{n} b_{n} e^{-E_{n}^{2}s^{2}/4} [1- i \cdot Erfi(\frac{E_{n} s}{2})] \ket{\psi_{n}},
    \end{aligned}
\end{equation}
where Erfi is imaginary error function. The Taylor-series expansion at $s \to \infty$ of above equation is
\begin{equation}
    \ket{\psi^{\prime}} = - i \sqrt{\frac{2}{\pi}} s^{-1} \sum_{n} ( E_{n}^{-1} + O(s^{-2}) ) \ket{\psi_{n}},
\end{equation}
where $\sqrt{\frac{2}{\pi}} s^{-1}$ is successful projection rate. Only in this case, the result is the desired state. And the error of final state is proportional to $s^{-2}$.

\section{Arbitrary qubit-qumode quantum gate construction}
\label{Arbitrary qubit-qumode local operator construction}

In this part, we will discuss how to construct arbitrary qubit-qumode unitary operator $e^{-i \hat{H}_{l} \hat{p}}$, where $\hat{H}_{l}$ is a tensor product of Pauli matrices and $\hat{p}$ is the momentum operator. Starting from one qubit situation, we firstly show that all the four Pauli matrices can be transformed from a Pauli X matrix. For convenience, we denote $X=\hat{\sigma}_{x}$, $Y=\hat{\sigma}_{y}$, $Z=\hat{\sigma}_{z}$, and $I=\hat{I}_{2}$.

\begin{equation}
    HXH=Z,\ SXS^{\dagger}=Y,\ XX=I
\end{equation}
The one-qubit-one-qumode unitary operator can be realized by the following quantum circuit.

\includegraphics[width=0.4\textwidth]{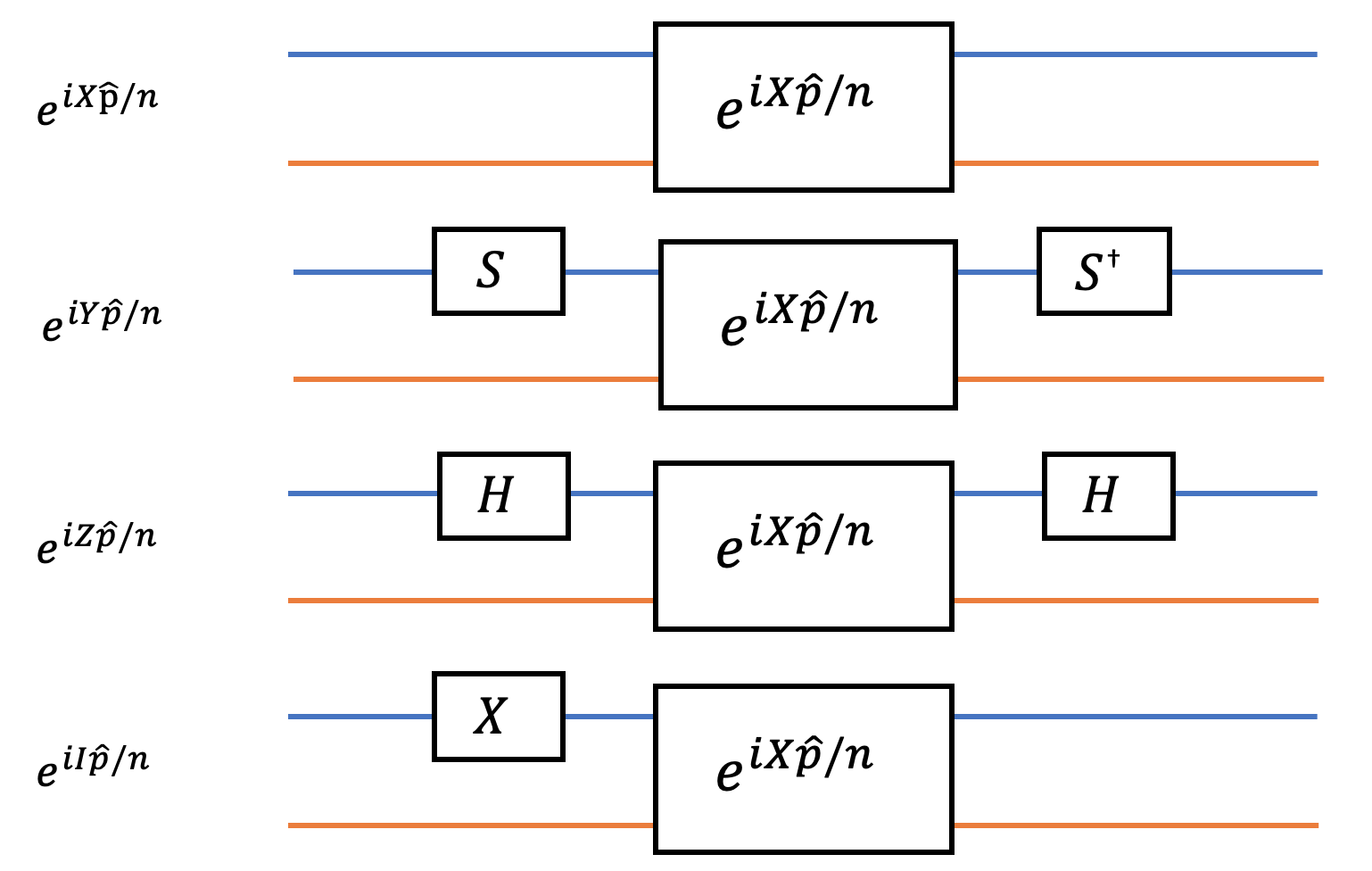}

For more than one qubits situation, we use Controlled-NOT (CNOT) gate to entangle the qubits.
\begin{equation}
    CNOT_{1,2} (X_{1} \otimes I_{2}) CNOT_{1,2}= X_{1} \otimes X_{2},
    CNOT
\end{equation}
where $CNOT_{1,2}$ is a CNOT gate controlled by the 1st qubit, targeting to the 2nd qubit.

So, two-qubits-one-qumode unitary operator $e^{iX_{1}X_{2}\hat{q}/n}$ can be implemented by the combination of $e^{iX\hat{q}/n}$ and a CNOT gate.

\includegraphics[width=0.4\textwidth]{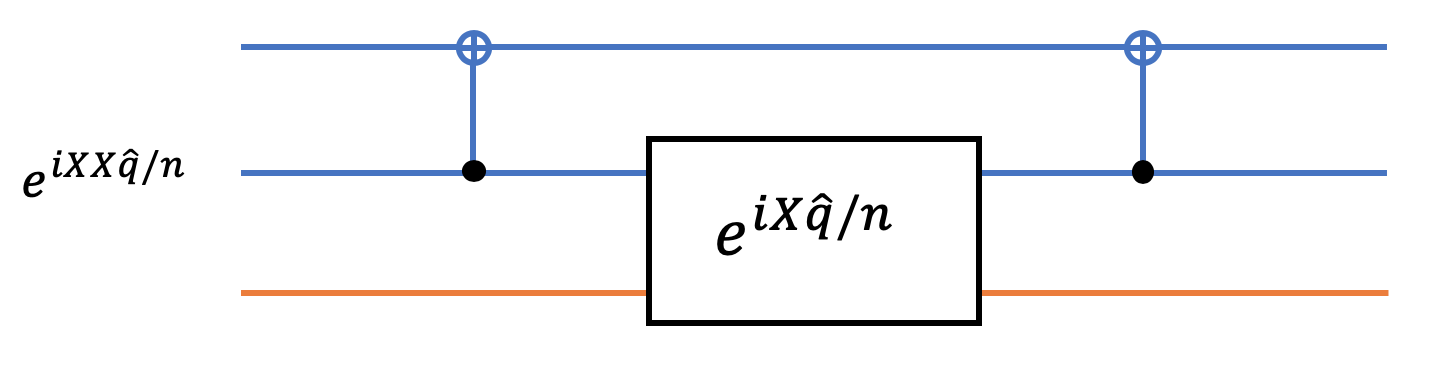}

By using the two-qubit CNOT gate that entangles the qubits, and single qubit gate that transforms the Pauli X operator to other Pauli operators, we can construct an evolution operator of tensor product of arbitrary Pauli operators $e^{-i\hat{H}_{l}\hat{p}}$. The following quantum circuit is a good example.

\includegraphics[width=0.4\textwidth]{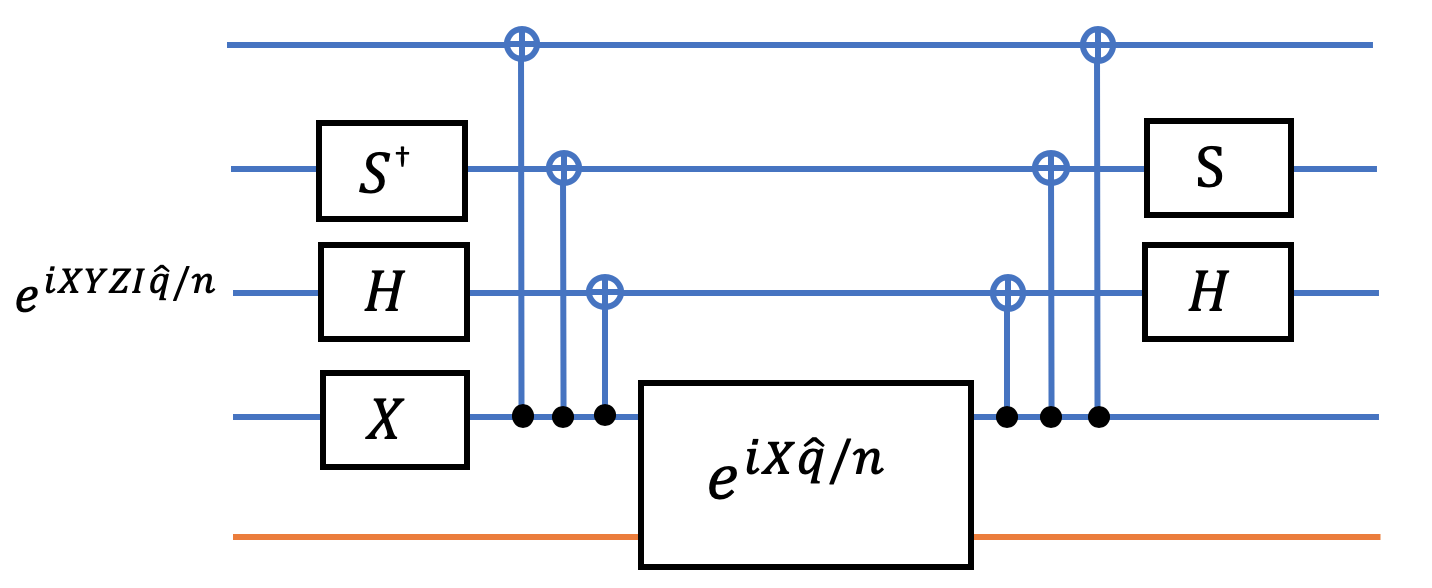}

\section{Hybrid quantum-classical method}
\label{Sec Hybrid quantum-classical method}
In this part, we propose a hybrid quantum-classical strategy of IPI that without any ancillae and thus postselection. The inverse Hamiltonian is still expressed  by linear combination of unitaries~\cite{kyriienko2020quantum}, but it is realized as a summation classically. Concretely, the inverse Hamiltonian is presented as a summation of evolution operators with different time, i.e., $\hat{H}^{-1} \approx \sum_{j=0}^{M_j-1} e^{-i\hat{H} j \Delta p} \Delta p$, where $\Delta p$ is the discrete interval of the summation, and $M_{j}$ is the up limit of the summation. The maximal evolution phase is defined as $\phi_{max} = M_{j} \Delta p$.
For higher order, the inverse Hamiltonian is approximated as
\begin{equation}
    \begin{aligned}
        \hat{H}^{-k} & \approx \sum_{j_{1}=0}^{M_{j_{1}}-1} \cdots \sum_{j_{k}=0}^{M_{j_{k}}-1} e^{-i\hat{H} j_{1} \Delta p} \cdots e^{-i\hat{H} j_{k} \Delta p} \cdot \Delta p^{k}\\ 
        & = \sum_{J} \hat{U}(J) \cdot \Delta p^{k}.
        \label{hybrid inverted Hamiltonian}
    \end{aligned}
\end{equation}

The target state is evolved to $\ket{\psi} = \hat{H}^{-k}\ket{b}  \approx \sum_{J} \hat{U}(J) \cdot \Delta p^{k} \ket{b}$. Then the ground state energy is expressed as
\begin{equation}
    \begin{aligned}
        E \approx \sum_{J} \sum_{J^{\prime}} \Delta p^{2k} \braket{b|\hat{U}(J)^{\dagger} H \hat{U}(J^{\prime})|b}.
        \label{hybrid energy}
    \end{aligned}
\end{equation}
Each expectation value can be parallelly computed by SWAP test~\cite{buhrman2001quantum} or QEE~\cite{peruzzo2014variational}, then summed up with corresponding weights to estimate the ground state energy.

\begin{figure}
    \centering
    \subfigure[]{
        \begin{minipage}[b]{0.45\textwidth}
            \includegraphics[width=1\textwidth]{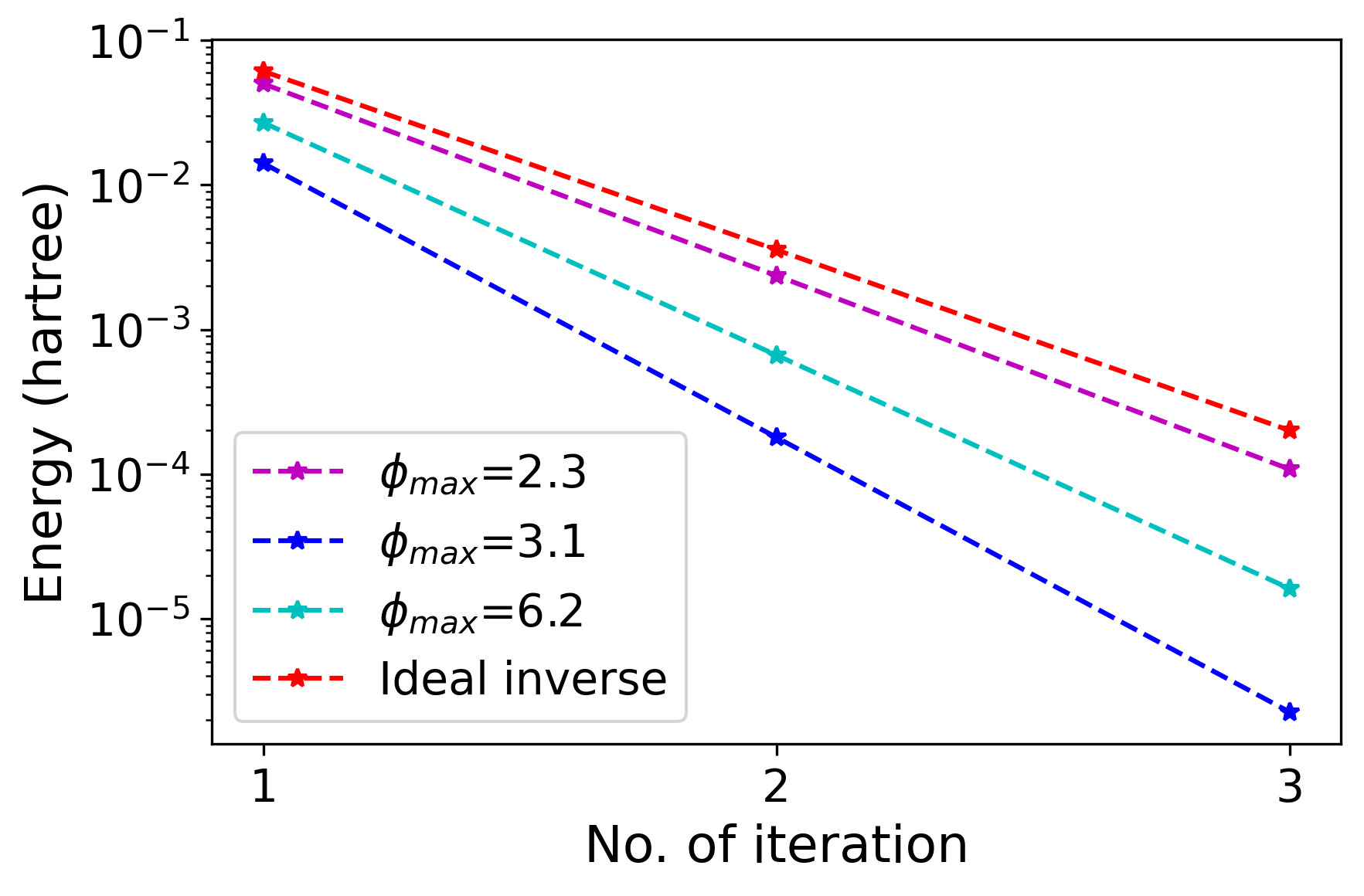}
        \end{minipage}
        \label{Hybrid H2 M}
    }
    \subfigure[]{
        \begin{minipage}[b]{0.45\textwidth}
            \includegraphics[width=1\textwidth]{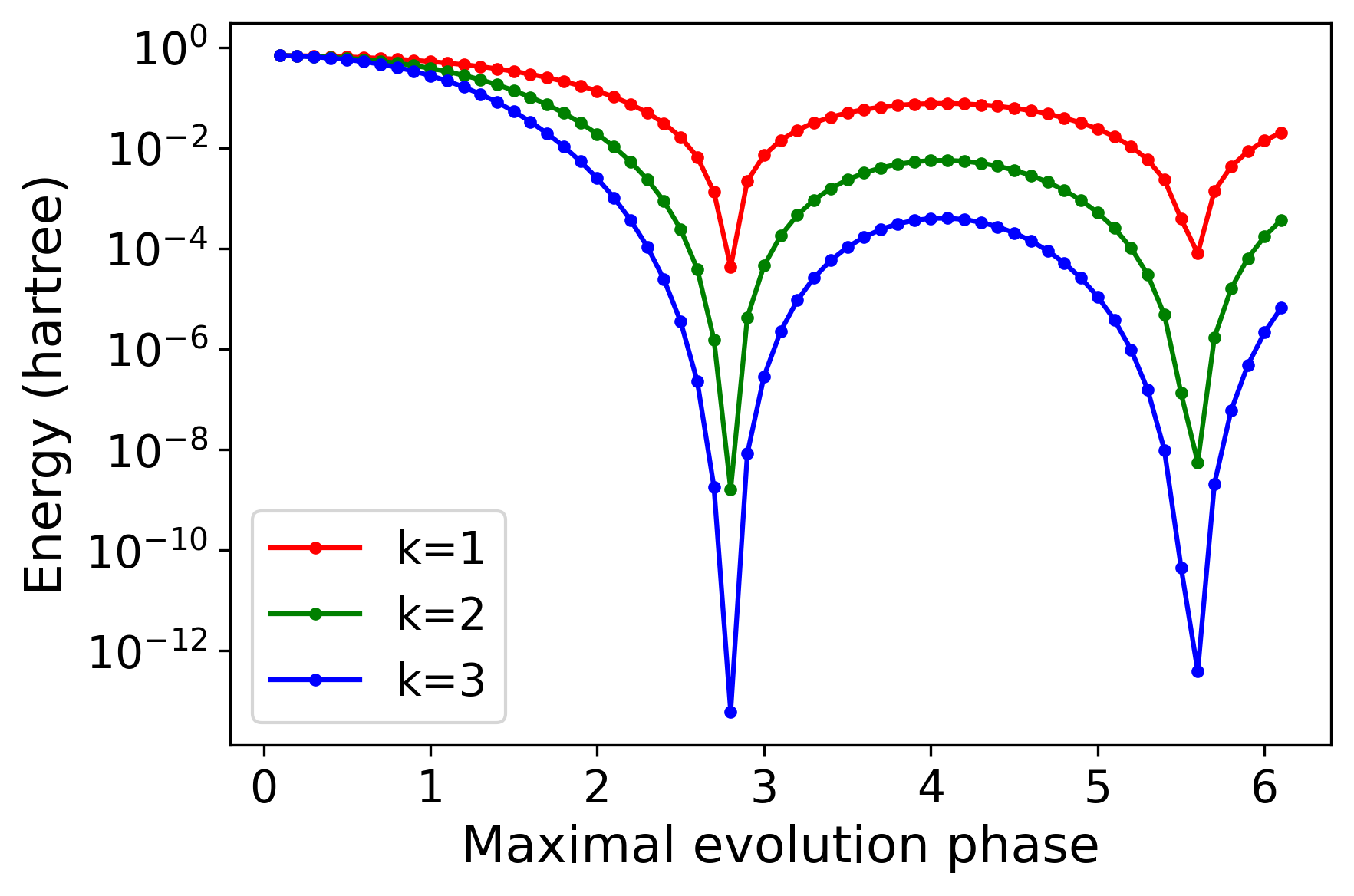}
        \end{minipage}
        \label{Hybrid H2 k}
    }
    \caption{(Color online)~Solving ground state energy of hydrogen molecular with the bond distance being equal to 0.75 by the hybrid quantum-classical algorithm. (a) For a chosen maximal evolution phase $\phi_{max}$, the relationship between the error $|E_{estimate}-E_{exact}|$ and number of iteration $k$. The red line is the result with using the ideal inverted Hamiltonian $H^{-1}$. (b) For a chosen number of iteration $k$, the relationship between the result error and maximal evolution phase $\phi_{max}$.} 
    \label{Hybrid figure}
\end{figure}

We simulate the process of this method to solve $H_{2}$ at the situation that bond distance equal 0.75$\AA$

and the discrete interval $\Delta p = 0.1$. For four chosen maximal evolution phase $\phi_{max}$, the relation between error and number of iteration $k$ is shown in Fig.~\ref{Hybrid H2 M}. For fixed number of iteration $k$, the energy difference at different maximal evolution phase $\phi_{max}$ is shown in Fig.~\ref{Hybrid H2 k}.

In this hybrid quantum-classical strategy, matrix inversion is realized by a series of unitary operators with different evolution time from $0$ to $(M_{j} - 1) \Delta p$. Considering Trotter decomposition, the number of gates required of unitary evolution increases with square of time~\cite{campbell2019random}. For this reason, long time evolution leads to high circuit coherent time requirement and long total runtime. By comparison, QuIPI does not have this concern, since the linear combination of unitaries in QuIPI is assisted by qumode resource state and an evolution of $\hat{H}\hat{p}$ with a fixed time period $t=1$.

\section*{References}
\bibliographystyle{iopart-num}
\bibliography{Reference}
\end{document}